\documentclass[10pt,aps,prd,twocolumn,showpacs,eqsecnum,superscriptaddress]{revtex4-1}
\usepackage[T1]{fontenc}
\usepackage[utf8]{inputenc}
\usepackage{lmodern}
\usepackage[english]{babel}
\usepackage{bm}
\usepackage{graphicx,amssymb,amsthm,amsmath}
\usepackage{hyperref}
\usepackage[export]{adjustbox}
\def \d {{\rm d}}

\begin{document}

\title{Null Points in the Magnetosphere of a Plunging Neutron Star}

\author{Ondřej Kopáček}
\email{kopacek@ig.cas.cz}
\affiliation{Astronomical Institute, Czech Academy of Sciences, Bo\v{c}n\'{\i} II 1401, Prague, CZ-141 31, Czech Republic}
\author{T. Tahamtan}
\email{tahamtan@utf.mff.cuni.cz}
\affiliation{Institute of Theoretical Physics, Faculty of Mathematics and Physics, Charles University, V~Hole\v{s}ovi\v{c}k\'ach 2, 180~00 Prague 8, Czech Republic}
\affiliation{Astronomical Institute, Czech Academy of Sciences, Bo\v{c}n\'{\i} II 1401, Prague, CZ-141 31, Czech Republic}
\author{Vladimír Karas}
\email{vladimir.karas@cuni.cz}
\affiliation{Astronomical Institute, Czech Academy of Sciences, Bo\v{c}n\'{\i} II 1401, Prague, CZ-141 31, Czech Republic}

\begin{abstract}
We explore the structure of a dipole-type vacuum field of a slowly 
rotating magnetic star near the horizon of a supermassive black hole, 
where the structure of field lines becomes highly distorted by effects 
of strong gravity. Such a situation may occur near a neutron star in the 
final stages of a plunging trajectory into a galactic center. We solve Maxwell's equations in the Rindler approximation for the rotating conducting source of dipolar magnetic field arbitrarily inclined with respect to the axis of rotation. For the fixed inclination angle we calculate the field including the radiative terms while in the general case we discuss the electromagnetic field considering the near-field terms only. In the latter case we investigate the emergence of magnetic null points within the vacuum magnetosphere. Null points become highly relevant in the presence of astrophysical plasma where they are connected with processes of magnetic reconnection and mass ejection.
\end{abstract}

\maketitle

\section{Introduction}
\label{intro}
Neutral points of the magnetic field are special locations where the magnetic intensity vanishes and the magnetic lines of force are bent around to create a critical (null) point. These magnetic nulls are an essential ingredient for the emergence of reconnection in magnetized plasmas. In classical (nonrelativistic) magnetohydrodynamics a complicated motion of the medium together with frozen-in magnetic lines is necessary. In strong gravity, however, even electrovacuum magnetic lines can become entangled and form null points. They reconnect once a conducting plasma is injected in the system or electron-positron pairs start to emerge. We study if the conditions for such magnetic nulls are met when a magnetized star, represented by a dipole, rotates and moves rapidly near the horizon of a supermassive black hole.

In general relativity, the lines of force of an electromagnetic field are
influenced by the gravitational field. Vice versa, the electromagnetic field stands as a source in coupled Einstein-Maxwell equations for the gravitational field and contributes
to the spacetime curvature. For astrophysical applications, however, a
limit of weak electromagnetic field is adequate to describe processes
near supermassive black holes (SMBH), such as the motion of magnetic
stars and the structure of magnetized accretion disks. Even the most
intense magnetic field of a magnetar can be treated quite accurately
within the test field approach \cite{beskin15}. Nonetheless, the
influence of the black hole strong gravity on the magnetic field lines
is important and it has to be taken into account; in particular, the
dipolar structure of a magnetized stellar body becomes highly
distorted due to the curved spacetime and fast motion.

Previously, the effects of relativistic
frame-dragging by which a rotating black hole acts on the magnetic lines
of force within and in the immediate vicinity of its ergosphere were studied in \cite{karas09,karas12}. While a number of simplifying assumptions have been imposed, namely, the case of a weak (test) magnetic field on the fixed (curved) background of a Kerr metric was analyzed, the authors could identify the main aspects leading to the creation of magnetic null points and the emergence of separatrices that mark the antiparallel lines of force {\em within} the electrovacuum spacetime. These crucial features are: (i)~rapid
rotation of the black hole (spin parameter $|a|\rightarrow1$ in
dimensionless geometric units); (ii)~translatory boost of the black hole (linear motion $v_{\rm t}\rightarrow1$); and (iii)~nonzero inclination of the asymptotic direction of the magnetic field with respect to the black hole rotation axis (oblique field with an off-axis component $B_\perp \neq0$).

We consider magnetic fields that are organized on large scales, i.e.,
the length scales exceeding the radius of the event horizon, $r\gtrsim
R_+=1+\sqrt{1-a^2}$ (measured in the units of a gravitational radius, $R_{\rm
g}=GM/c^2\sim1.5\times10^{5}M_{\bullet}/M_{\odot}\;$cm). An
asymptotically homogeneous magnetic field in orientation perpendicular
to the rotation axis is probably the simplest nontrivial example of
such an ordered field structure. However, it may not be the most
relevant example from the viewpoint of astrophysical applications; the
uniform magnetic field assumes that the source currents are of
external origin with respect to the black hole and that they flow very far
in spatial infinity, $r\gg R_{\rm g}$. Strictly speaking this
constraint violates the assumption of a test magnetic field because
the total energy stored in the asymptotically uniform field leads to
the spacetime that does not satisfy the asymptotic flatness; instead,
it resembles the cylindrical universe with the symmetry dictated by
the field line direction \cite{melwin64,ernst76,Tahamtan-Melvin}.

The choice of electrovacuum spacetime imposes limitations on the
astrophysical relevance of the adopted approach. On the other hand,
such a solution represents a clean system and it allows us to reveal
the geometrical effects of strong gravity, which could be otherwise
hidden if we attempt to explore more realistic (complicated)
situations. The assumptions given above imply that we do not consider
the effects of small-scale turbulent magnetic fields that can develop
within the plasma due to magnetorotational instability \cite{balbus91} and/or powerful shearing motions \cite{koide08, asenjo17}.

Although the black hole frame-dragging twists the lines of magnetic
force to such an extreme extent that X-shaped critical null points
develop by a purely geometrical effect, these field lines would not be
able to reconnect unless some dissipative plasma fills the regions \cite{priest00,somov06}. In fact, it appears rather
inevitable that a black hole magnetosphere is continuously filled by
electron-positron plasma due to pair creation, or the gaseous material
can be brought into the region by transiting stars and stellar
remnants. Purely vacuum solutions serve as a starting configuration
and test bed that must be inevitably modified by plasma in an
astrophysically realistic environment for instance, when plasma of a
pulsar wind nebula is injected into the SMBH neighborhood of a galactic
core. Turbulent motions can even enhance the reconnection rate
\cite{lazarian15} but these effects are beyond the scope of the
present paper.

The emergence of magnetic neutral points appears to be a generic
property that occurs near the ergosphere even if the magnetic
structure happens to be different from the asymptotically uniform
condition, although the shape of the field lines is progressively departing from this simple solution as one proceeds to
the length scales of order $\sim R_{\rm g}$ and more. In particular,
we want to examine the conditions for reconnection of the field lines
associated with a magnetic star on a close trajectory around the black
hole. Such a setup is relevant, for example, to describe the
interaction of a magnetar in a galactic center nuclear star cluster.

A dipole-type magnetic field is obviously a better approximation to
describe a magnetic star near SMBH, and we embark on its exploration
in the present paper. In analogy with our previous results \cite{karas09,karas12} we expect that the boost and rotation (of both the
supermassive black hole and the massive magnetic dipole representing
the plunging star) will play an important role on the condition for
the magnetic null points, and we need to explore the case of a close
approach (a plunging trajectory). To simplify the mathematical
formalism while preserving the essential properties of the physical
system we adopt the framework of Rindler approximation \cite{rindler66,
macdonald85, orazio13}. Rindler geometry is a
geometrically simpler representation of the Schwarzschild black hole in
the near-horizon regime. This allows us to model the
(electro)magnetic structure of field lines of a magnetic star near
the black hole horizon as an accelerated dipole in the Minkowski
spacetime.

Rindler spacetime is flat and its metric can be written in Minkowski coordinates $(T,X,Y,Z)$:
\begin{eqnarray}
\label{mink} 
ds^2 & = & -dT^2 + dX^2 + dY^2 + dZ^2 \\
\label{rindler}
 & = & -\alpha^2 dt^2 + dx^2 + dy^2 +dz^2,
\end{eqnarray}
where that lapse function $\alpha$ connects the Rindler coordinates $(t,x,y,z)$ and the corresponding approximation of the spatial part of the Schwarzchild metric near the horizon \cite{macdonald85},
\begin{eqnarray}
ds^2 & = & \left(1-2M_\bullet/r\right)^{-1}\,dr^2 + r^2\,d\Omega^2\\
 & \simeq & g_h^{-2}d\alpha^2 + R_{\rm G}^2\,d\Omega^2
\end{eqnarray}
where $g_h\simeq \alpha/z$ denotes the horizon surface gravity and $z$ is the proper distance from the horizon which is related to the Schwarzschild radial coordinate $r$ as $z\simeq 4M_\bullet\left(1-2M_\bullet/r\right)^{1/2}$. The Rindler approximation thus neglects the spatial curvature near the horizon and it approximates the black hole gravity solely by the corresponding gravitational acceleration. Therefore, the Rindler representation neglects all effects related to the spacetime curvature; nevertheless, it is a useful tool to capture the leading effects on the electromagnetic field structure very close to the horizon, where the acceleration plays the dominant role.

In this paper we employ the Rindler approximation in order to analyze the field of a magnetic dipole in various states of motion close to the Schwarzschild black hole. A magnetic dipole in Rindler spacetime can simulate a neutron star or magnetar in the vicinity of a supermassive black hole. For such a situation the neutron star can be considered almost pointlike while the black hole horizon is almost planar. In this setup we calculate the electromagnetic field including the radiative terms (Sec.~\ref{fields}). Further in Sec.~\ref{magnetosphere}, we consider a more realistic model of a neutron star of finite size and arbitrary inclination of the dipole field with respect to the rotation axis. In Sec.~\ref{nps} we investigate the presence and location of the null points of the magnetic field and discuss the role of parameters of the model. Results of the analysis are concluded in Sec.~\ref{conclusions}. 

\section{Rotating magnetic dipole in Minkowski and Rindler spacetimes}
\label{fields}
Using the Lorentz gauge, the Maxwell's equations  for the 4-potential $A^{\alpha}$ in Minkowski spacetime (\ref{mink}) are given as
\begin{equation}
\label{maxwell}
\Box A^{\alpha}=\frac{4\,\pi}{c}\,J^{\alpha},
\end{equation}
where the left-hand side contains the D'Alembert operator $\Box$ in the flat spacetime, and the right-hand side contains the source term of the electromagnetic field. Gaussian units are used in (\ref{maxwell}); however, for the rest of paper we employ geometric units setting $G=c=1$.

The 4-current along the dipole worldline is
\begin{equation}
J^{\alpha}=\nabla_{\mu}\int{Q^{\alpha\,\mu}(\tau)\delta^{(4)}[x_{o}-x_{s}(\tau)]\,\d \tau}
\end{equation}
where $\tau$ is the proper time of the dipole source with spacetime coordinates $x_{s}$ while the observer is located at $x_{o}$. The antisymmetric dipole tensor, 
\begin{equation}\label{Charge formula}
Q^{\alpha\,\mu}(\tau)=V^{\alpha}\,p^{\mu}-p^{\alpha}\,V^{\mu}+\epsilon^{\alpha\,\mu}{}_{\rho\,\sigma}V^{\rho}\,m^{\sigma},
\end{equation}
is composed of an electric part $p^{\mu}$ and magnetic part $m^{\mu}$, $V^{\mu}$ is the 4-velocity of the source and $\epsilon$ is the Levi-Civit\`{a} symbol.

\begin{figure*}[htb]
\center
\includegraphics[scale=.48]{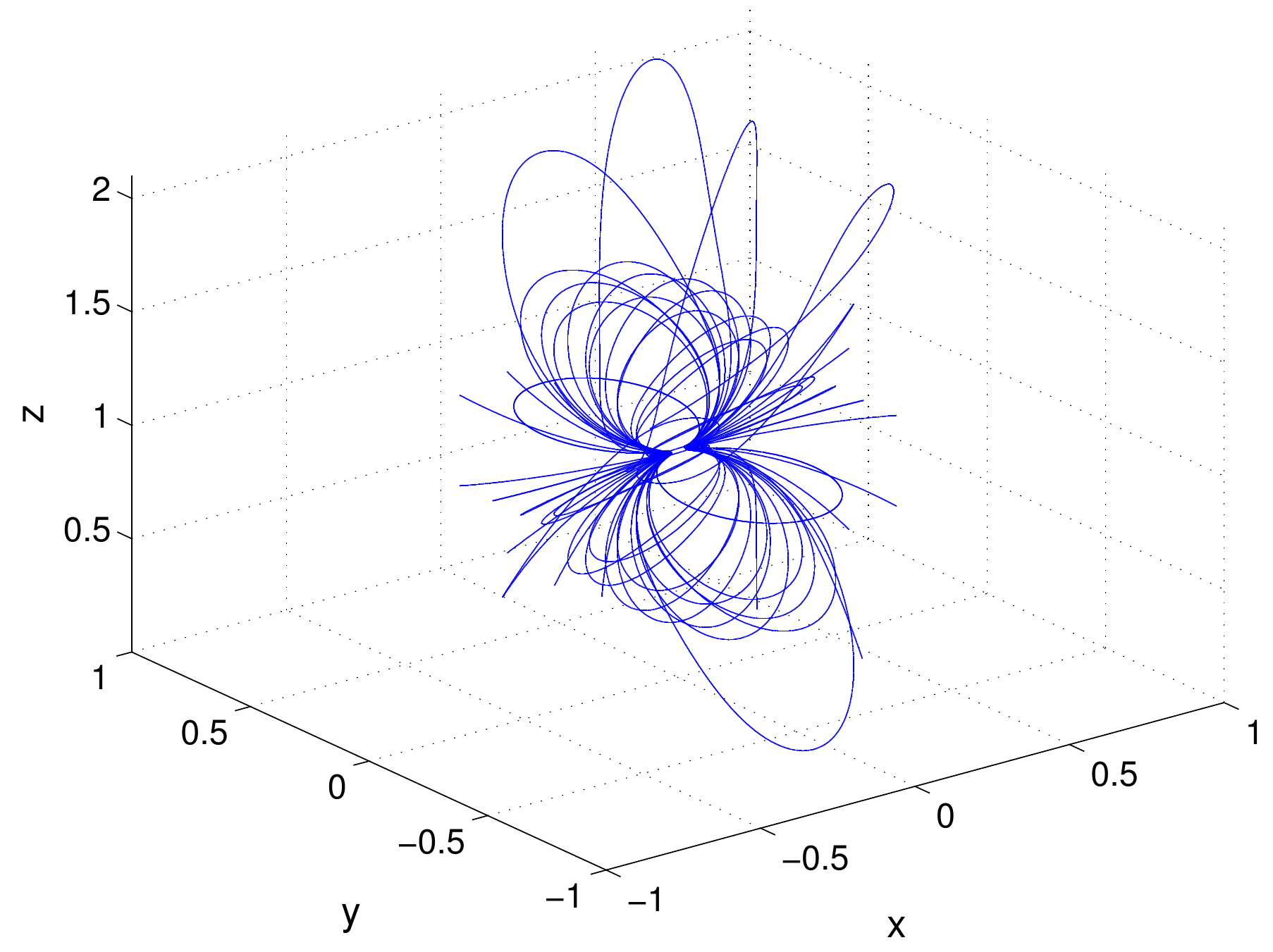}
\includegraphics[scale=.48]{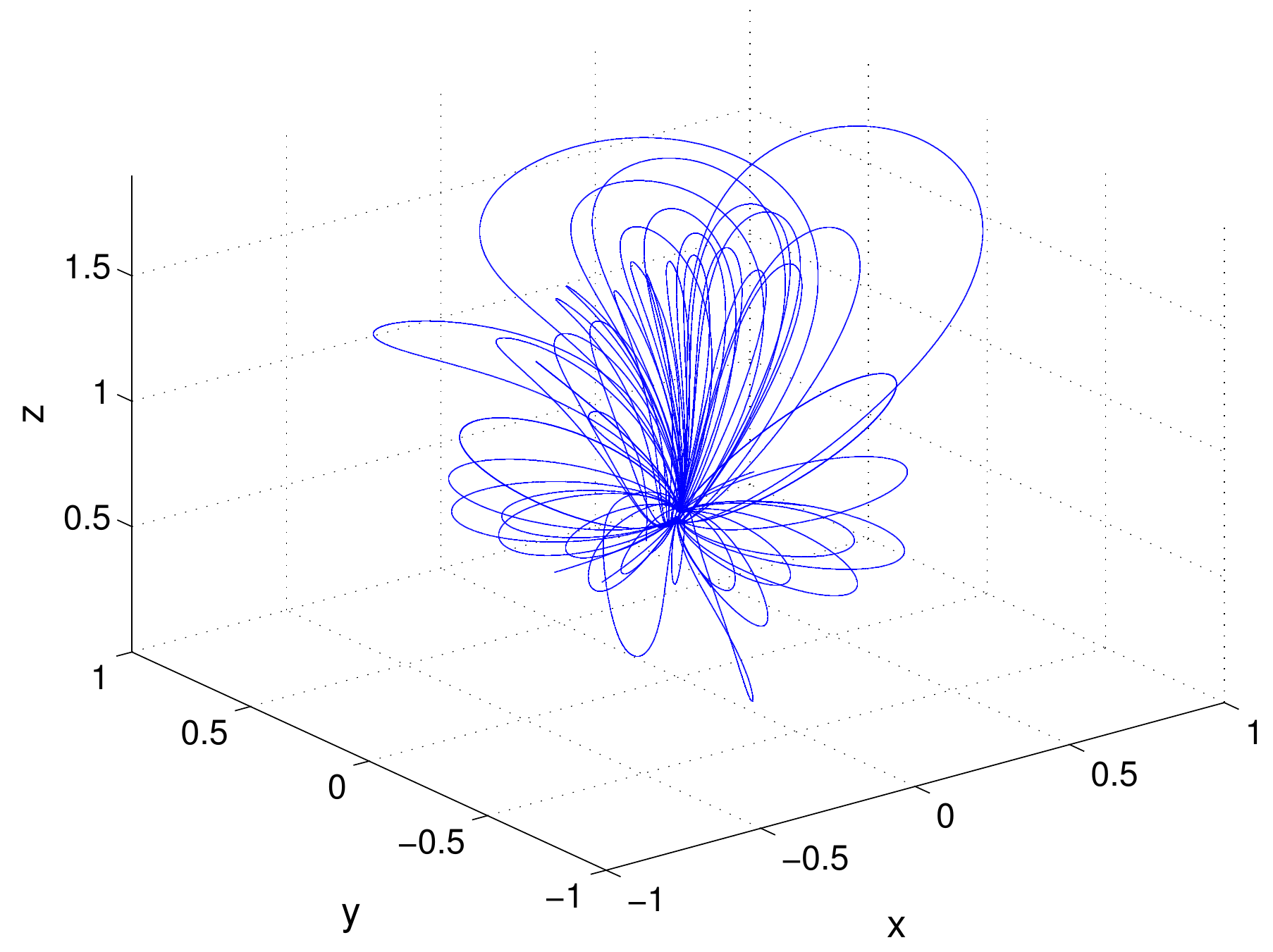}
\caption{Comparison of the magnetic field lines of a rotating dipole ($\omega=1$, $m=1$) as observed in the inertial reference frame ($g_H t=0$, left panel) and in the accelerated Rindler frame ($g_H t=1$, right panel). The dipole is located at $Z_s=1$.}
\label{3D_radiative}
\end{figure*}

The solution for $A^{\alpha}$ can be written as \cite{orazio13}
\begin{equation} \label{vector potential}
A^{\alpha}=\nabla_{\mu}\left[\frac{Q^{\alpha\,\mu}}{g_{\beta\gamma}r^{\beta}V^{\gamma}}\right]_*,
\end{equation}
where the subscript $*$ denotes the evaluation at retarded time. Since the source is moving, the observer with the position 3-vector $\bm{X}$ at time $T$ will observe fields generated by the source in the past. Therefore, we evaluate all quantities at the retarded time. The retarded time is found as a function of observer coordinates by imposing the null condition. The relative distance between an observer and a point on the trajectory of the source $x_s=\{T_{s},\bm{X}_s\}$ is given as
\begin{eqnarray}
r^{\mu}&=&\{r^0,\bm{r}\} \nonumber \\
&=&\{T-T_{s},\bm{X}-\bm{X}_{s}(T_{s})\}.
\end{eqnarray}
The null condition $(r_{\mu}r^{\mu})_{*}=0$ at retarded time leads to
\begin{equation}\label{null condition}
(T-T_{*})^2=(\bm{X}-\bm{X}_{s}(T_{*}))^2.
\end{equation}
Electromagnetic tensor $F_{\alpha\beta}$ is given as 
\begin{equation}
F_{\alpha\beta}=\nabla_{\alpha}A_{\beta}-\nabla_{\beta}A_{\alpha},
\end{equation}
and electric a magnetic fields as measured by the observer with 4-velocity $u^{\mu}$ are
\begin{equation}
E^{\alpha}=F^{\alpha\,\mu}u_{\mu}\,, \qquad\qquad{} B^{\alpha}=\frac{1}{2}\epsilon^{\alpha\mu\gamma\delta}F_{\gamma\delta}\,u_{\mu}.
\end{equation}

\subsection{Electromagnetic Fields in Minkowski Spacetime}
We consider a constant magnetic dipole embedded in the $(X,Y)$-plane which is at rest ($V^{\alpha}=\{V^{0},0,0,0\}$) and rotates along the $Z$-axis,
\begin{equation}
m^{\mu}=\{0,m^{x},m^{y},0\},
\end{equation}
in which 
\begin{equation}
m_{x}=m\,\cos(\omega\,\tau) \qquad \qquad m_{y}=m\,\sin(\omega\,\tau),
\end{equation}
with $\omega$ denoting the angular velocity of magnetic dipole and $m$ its constant magnitude. Using the above assumptions, the 4-potential $A^{\alpha}$ may be expressed as \cite{orazio13}
\begin{equation}
A^{\alpha}=\left. \epsilon^{\alpha}{}_{\mu \rho \sigma}r^{\mu}\left[\frac{V^{\rho}\,\dot{m}^{\sigma}}{(r.V)^2}-\frac{V^{\rho}\,m^{\sigma}}{(r.V)^3}\right]\right\rvert_{*},
\end{equation}
where the ``dot'' denotes the $\tau$ derivative. The electromagnetic field tensor is then

\begin{widetext}
\begin{eqnarray}\label{fields-Min}
F_{\alpha \beta}&=&-\epsilon_{[\alpha \beta]\rho \sigma}\frac{V^{\rho}\,\dot{m}^{\sigma}}{(r.V)^2} +\epsilon_{[\alpha \beta]\rho \sigma}\frac{V^{\rho}\,m^{\sigma}}{(r.V)^3}
+\frac{r_{\left[\alpha \right.}\epsilon_{\left. \beta \right]\mu \rho \sigma}}{(r.V)^3}\{r^{\mu}\,V^{\rho}\,\ddot{m}^{\sigma}-V^{\mu}\,V^{\rho}\,\dot{m}^{\sigma}\}\nonumber \\
&&-\frac{r_{\left[\alpha \right.}\epsilon_{\left. \beta \right]\mu \rho \sigma}}{(r.V)^4}\{3\,r^{\mu}\,V^{\rho}\,\dot{m}^{\sigma}-V^{\mu}\,V^{\rho}\,m^{\sigma}\} 
+3\frac{V_{\left[\alpha \right.}\epsilon_{\left. \beta \right]\mu \rho \sigma}}{(r.V)^4}r^{\mu}\,V^{\rho}\,m^{\sigma}+3\frac{r_{\left[\alpha \right.}\epsilon_{\left. \beta  \right]\mu \rho \sigma}}{(r.V)^5}r^{\mu}\,V^{\rho}\,m^{\sigma}
-2\frac{V_{\left[\alpha \right.}\epsilon_{\left. \beta \right]\mu \rho \sigma}}{(r.V)^3}r^{\mu}\,V^{\rho}\,\dot{m}^{\sigma}. \nonumber\\
\end{eqnarray}

\begin{figure*}[htb]
\center
		\includegraphics[trim={2cm 1cm 3cm 1cm},clip,scale=0.22]{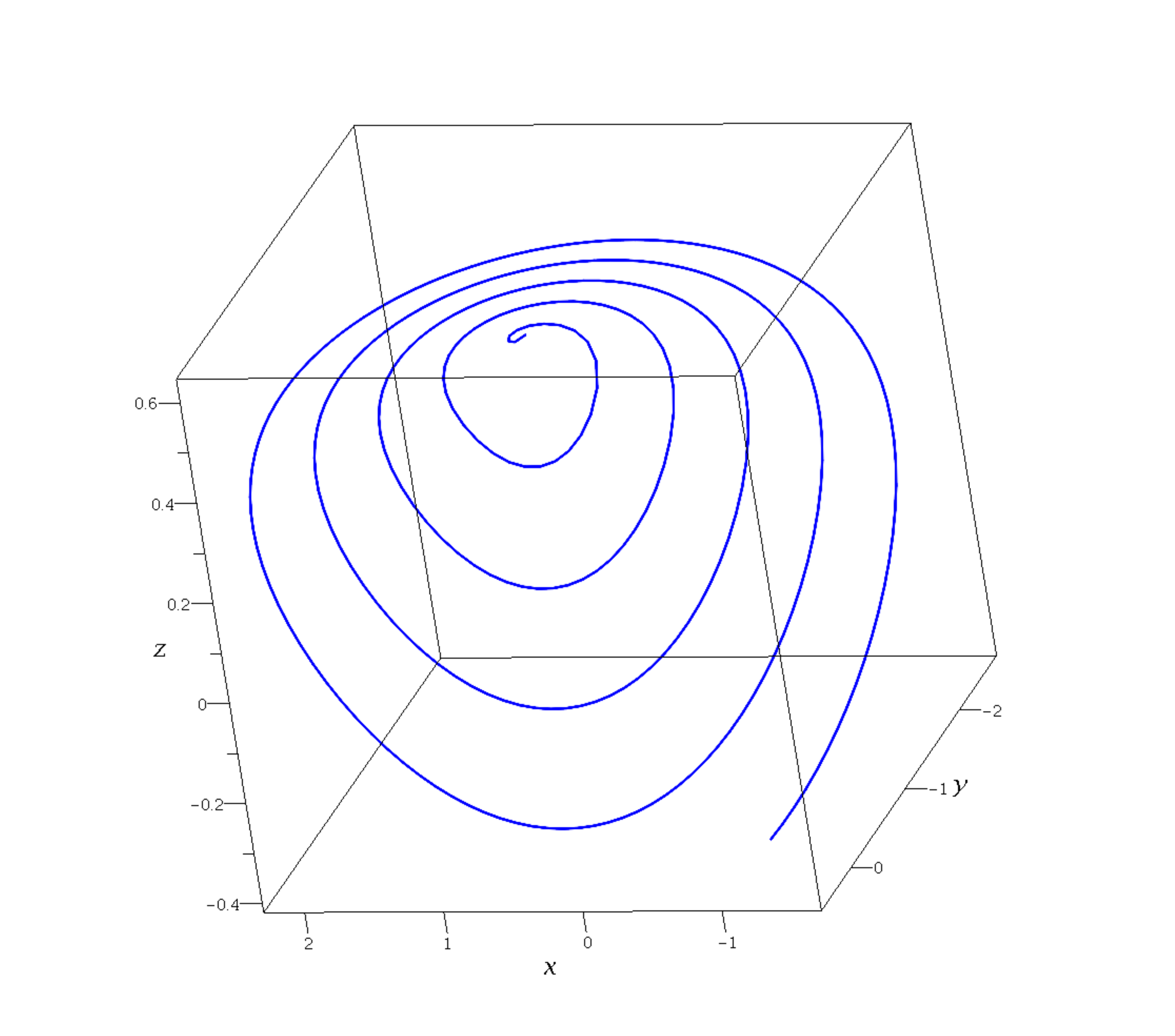}
		\includegraphics[trim={.5cm 1.5cm 1cm 1.7cm},clip,scale=0.29]{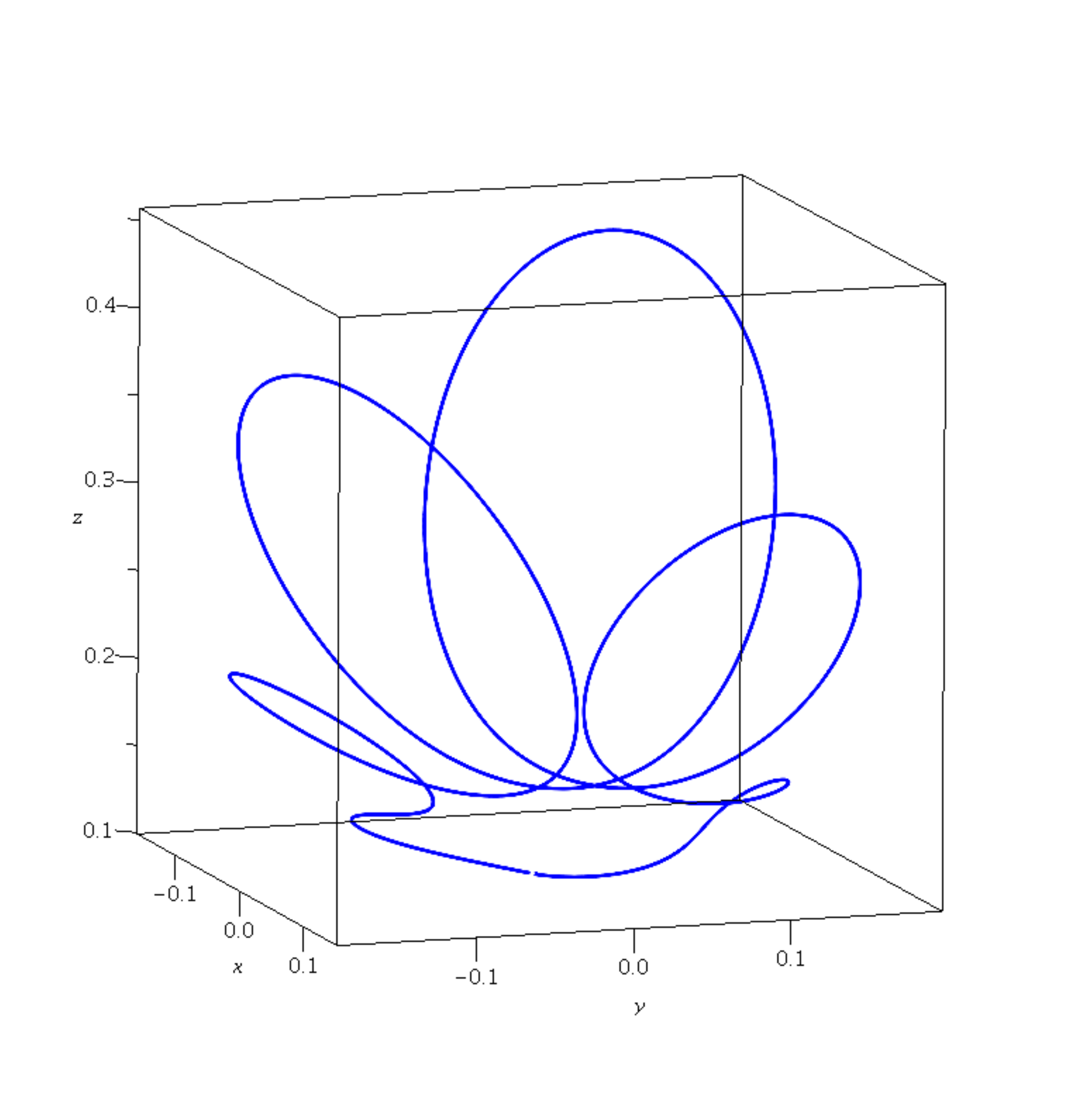}
			\caption{Comparison of two particular magnetic field lines in Rindler spacetime for the early time $(s =-1.6891785..30$, left panel) and for the late time $(s=-1.871..1.5876863$, right panel). Other parameters are set as $\omega=10$, $m=1$, $Z_{s}=1$ and $g_{H}=1$. The field lines are constructed with initial values  $x_{0}=0.1$, $y_{0}=0.1$ and $z_{0}=0.4$.	\label{Rindler}}
\end{figure*}

All the calculations are done at retarded time. We consider the static location of the magnetic dipole on the $Z$-axis at $\bm{X}_s=(0,0,Z{s})$. From (\ref{null condition}) we obtain
\begin{equation}\label{retarted equa}
T_{*}=T-r\,,\qquad r=\sqrt{X^2+Y^2+(Z-Z{s})^2},
\end{equation}
where $r$ hereafter denotes the distance from the dipole (not the Schwarzschild radial coordinate). From (\ref{fields-Min}) we find the electric and magnetic fields. The magnetic field components are 

\begin{eqnarray}\label{Bx-Mink}
B_{X}&=&F_{YZ}\rvert_{*} =\frac{2\,\left(r\,\dot{m}_{x}+{m}_{x}\right)}{r^3}-\frac{\left(r^2-X^2\right)\left(r^2\,\ddot{m}_{x}+3\,r\,\dot{m}_{x}+3\,{m}_{x}\right)}{r^5} 
+\frac{X\,Y\,\left(r^2\,\ddot{m}_{y}+3\,r\,\dot{m}_{y}+3\,{m}_{y}\right)}{r^5}\bigg|_{*}, \\ 
\label{By-Mink}B_{Y}&=&F_{ZX}\rvert_{*} =\frac{2\,\left(r\,\dot{m}_{y}+{m}_{y}\right)}{r^3}-\frac{\left(r^2-Y^2\right)\left(r^2\,\ddot{m}_{y}+3\,r\,\dot{m}_{y}+3\,{m}_{y}\right)}{r^5} 
+\frac{X\,Y\,\left(r^2\,\ddot{m}_{x}+3\,r\,\dot{m}_{x}+3\,{m}_{x}\right)}{r^5} \bigg|_{*}, \\
\label{Bz-Mink}B_{Z}&=&F_{XY}\rvert_{*} =\frac{\left(Z-Z{s}\right)}{r^5}\left\{r^2\,\left(X\,\ddot{m}_{x}+Y\,\ddot{m}_{y}\right)+3\,\left(X\,{m}_{x}+Y\,{m}_{y}\right)\right.
\left.+3\,r\,\left(X\,\dot{m}_{x}+Y\,\dot{m}_{y}\right)\right\} \bigg|_{*}.
\end{eqnarray}

The electric field components are
\begin{eqnarray}\label{Ex-Mink}
E_{X}&=&F_{XT}\rvert_{*} =-\frac{\left(Z-Z{s}\right)\left(\dot{m}_{y}+r\,\ddot{m}_{y}\right)}{r^3} \bigg|_{*},  \\
\label{Ey-Mink}E_{Y}&=&F_{YT}\rvert_{*} =\frac{\left(Z-Z{s}\right)\left(\dot{m}_{x}+r\,\ddot{m}_{x}\right)}{r^3}\bigg|_{*}, \\
\label{Ez-Mink}E_{Z}&=&F_{ZT}\rvert_{*} =\frac{Y\,\dot{m}_{x}-X\,\dot{m}_{y}-r\,\left(Y\,\ddot{m}_{x}-X\,\ddot{m}_{y}\right)}{r^3}\bigg|_{*}.
\end{eqnarray}
\end{widetext}

\subsection{Electromagnetic Fields in Rindler Spacetime}
In the previous section we found the electromagnetic fields in Minkowski spacetime. Since the Rindler observer represents a particular worldline in the flat spacetime, we can use the formulas obtained for Minkowski and apply the coordinate transformation to convert them into Rindler spacetime.

With the following coordinate transformation, 
\begin{equation}\label{Rindler transf}
T=z\,\sinh(g_{H}t),\,\,\,X=x,\,\,Y=y,\,\,\,Z=z\,\cosh(g_{H}t),
\end{equation} 
applied in (\ref{mink}) we indeed obtain the Rindler line element (\ref{rindler}) with the lapse function given as $\alpha=g_{H}z$. The electromagnetic fields in Rindler spacetime (primed) are thus obtained from the corresponding ones in Minkowski spacetime (unprimed) as follows:
\begin{equation}\label{field-transf}
E^{\alpha'}=\frac{\partial{x^{\alpha'}}}{\partial{x^{\alpha}}}F^{\alpha\,\mu}u^{R}_{\mu}\,, \qquad B^{\alpha'}=\frac{1}{2}\frac{\partial{x^{\alpha'}}}{\partial{x^{\alpha}}}\epsilon^{\alpha\mu\gamma\delta}F_{\gamma\delta}\,u^{R}_{\mu},
\end{equation}
where $u^{\mu}_{R}$ denotes the 4-velocity of the Rindler frame with respect to the Minkowski observer
\begin{equation}\label{Rindler velocity}
u^{\mu}_{R}=\{\cosh(g_{H}\,t),0,0,\sinh(g_{H}\,t)\}.
\end{equation}
The resulting electric field is 
\begin{widetext}
\begin{eqnarray}
\label{Ex}E_{x}&=&-\sinh(g_{H}\,t)\left[\frac{2\,\left(r\,\dot{m}_{y}+{m}_{y}\right)}{r^3}
-\frac{\left(r^2-y^2\right)\left(r^2\,\ddot{m}_{y}+3\,r\,\dot{m}_{y}+3\,{m}_{y}\right)}{r^5}  
+\frac{x\,y\,\left(r^2\,\ddot{m}_{x}+3\,r\,\dot{m}_{x}+3\,{m}_{x}\right)}{r^5}\right]\nonumber \\
&&-\cosh(g_{H}\,t) \left[\frac{\left(z\,\cosh(g_{H}t)-Z{s}\right)\left(\dot{m}_{y}+r\,\ddot{m}_{y}\right)}{r^3}\right], \\
\label{Ey}E_{y}&=&\sinh(g_{H}\,t)\left[\frac{2\,\left(r\,\dot{m}_{x}+{m}_{x}\right)}{r^3}-\frac{\left(r^2-x^2\right)\left(r^2\,\ddot{m}_{x}+3\,r\,\dot{m}_{x}+3\,{m}_{x}\right)}{r^5} 
+\frac{x\,y\,\left(r^2\,\ddot{m}_{y}+3\,r\,\dot{m}_{y}+3\,{m}_{y}\right)}{r^5} \right]\nonumber\\
&&+\cosh(g_{H}\,t)\left[\frac{\left(z\,\cosh(g_{H}t)-Z{s}\right)\left(\dot{m}_{x}+r\,\ddot{m}_{x}\right)}{r^3}\right], \\
\label{Ez}E_{z}&=&\frac{y\,\dot{m}_{x}-x\,\dot{m}_{y}-r\,\left(y\,\ddot{m}_{x}-x\,\ddot{m}_{y}\right)}{r^3}. \nonumber\\
\end{eqnarray}

\begin{figure*}[htb]
\center
\includegraphics[scale=.49]{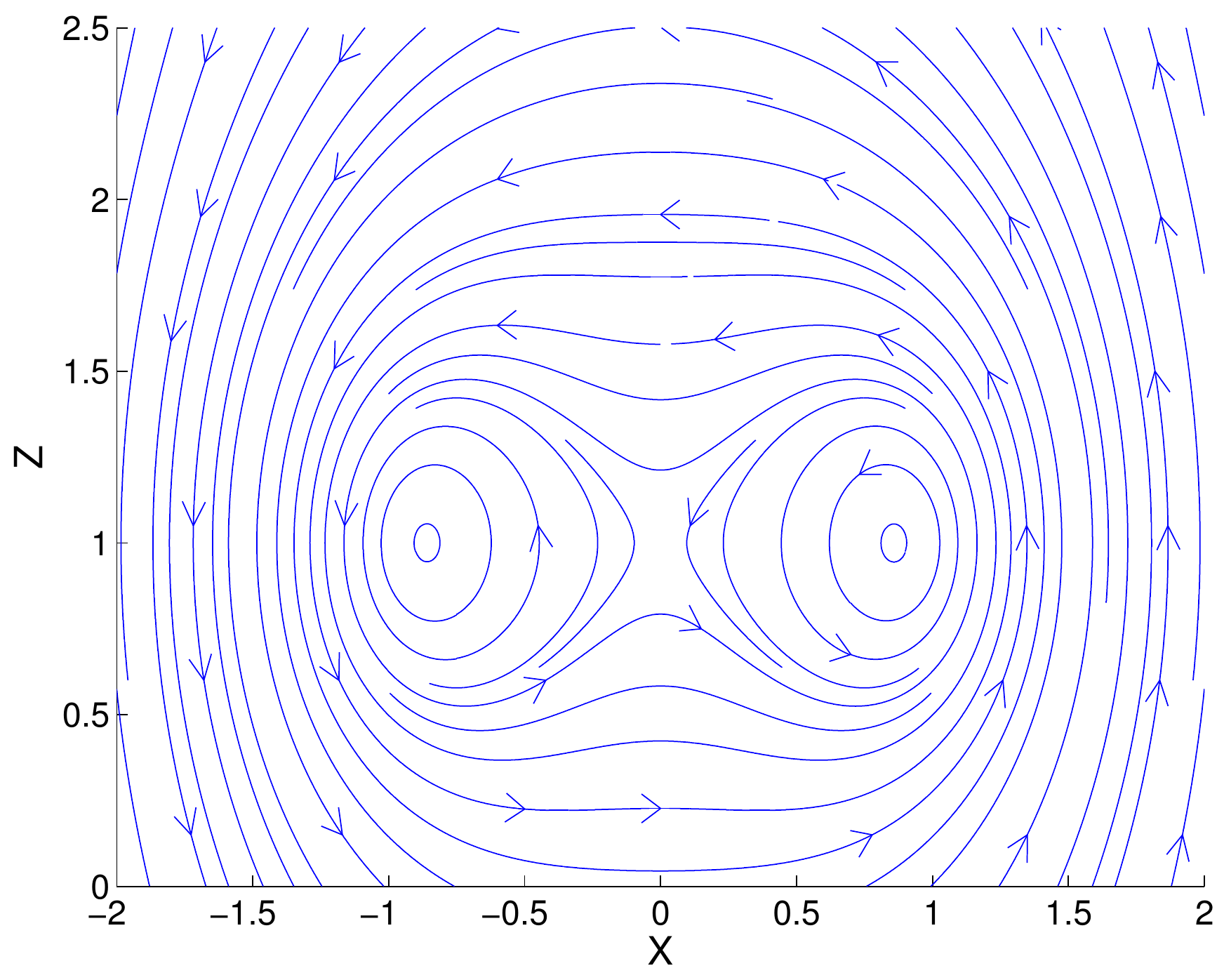}
\includegraphics[scale=.49]{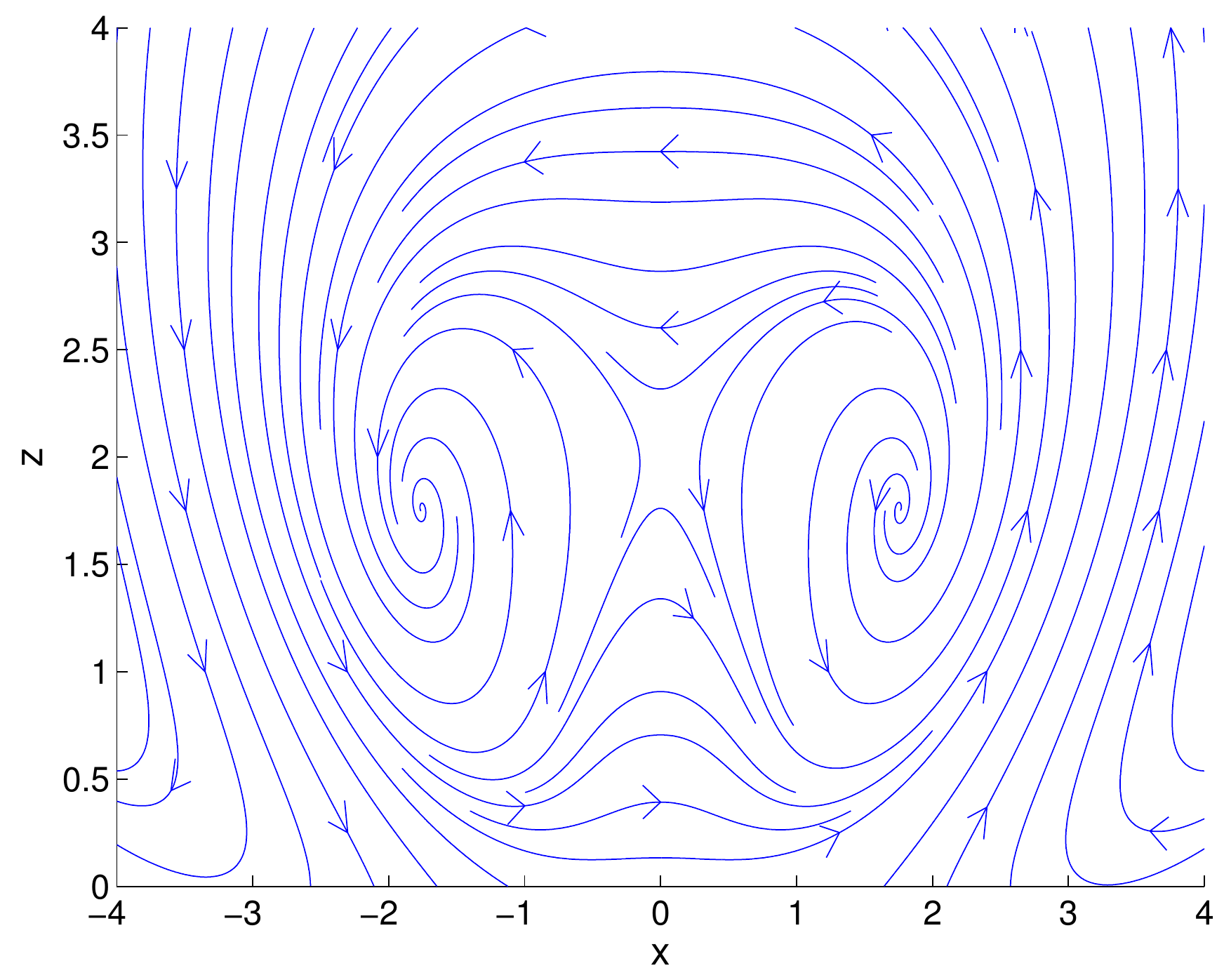}
\caption{Comparison of the electric field lines of a rotating dipole ($\omega=1$, $m=1$) as observed in the inertial reference frame ($g_H t=0$, left panel) and in the accelerated Rindler frame ($g_H t=1$, right panel). The dipole is located at $Z_s=1$.}
\label{el_radiative}
\end{figure*} 

\begin{figure*}[htb]
\center
\includegraphics[scale=.49]{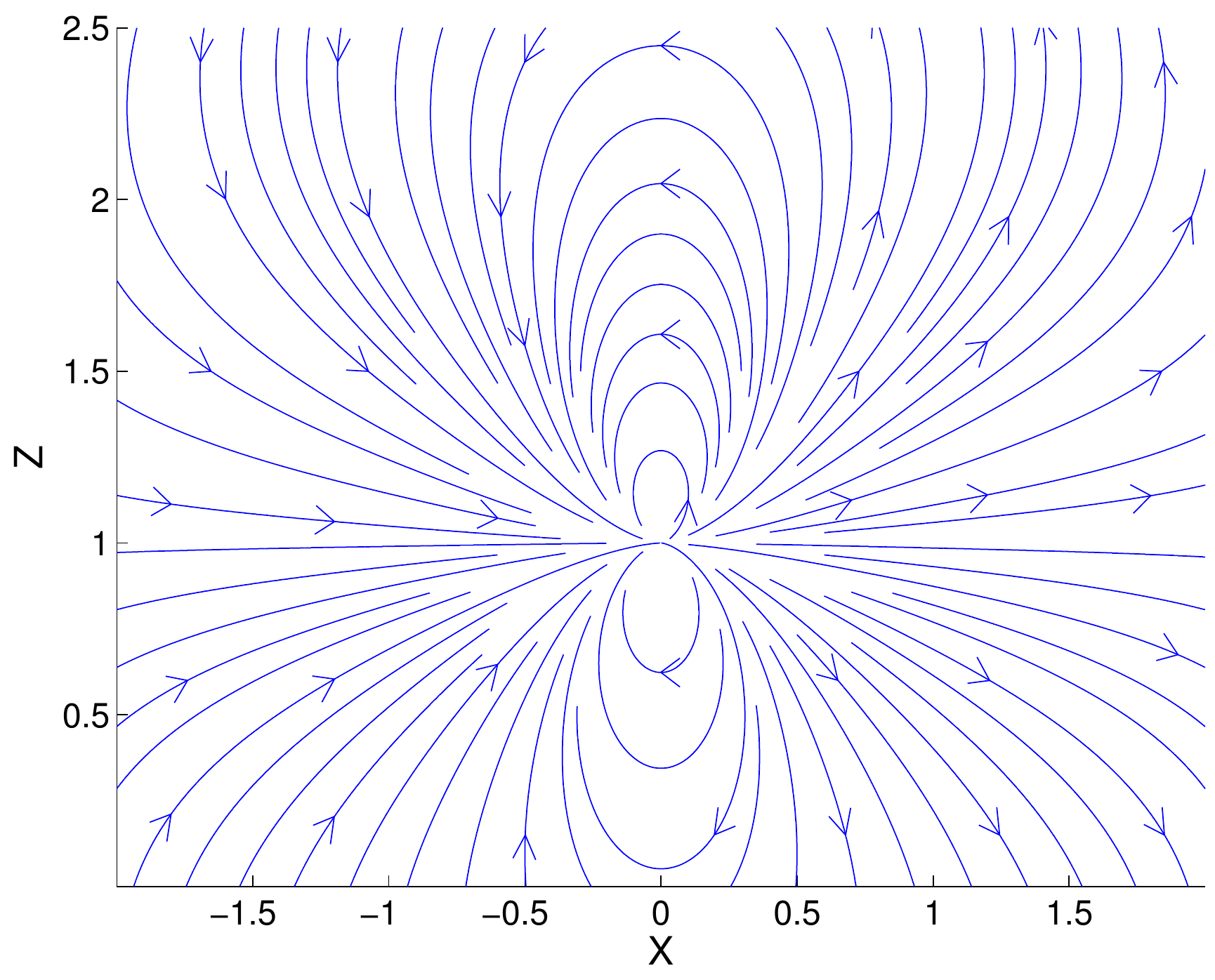}
\includegraphics[scale=.49]{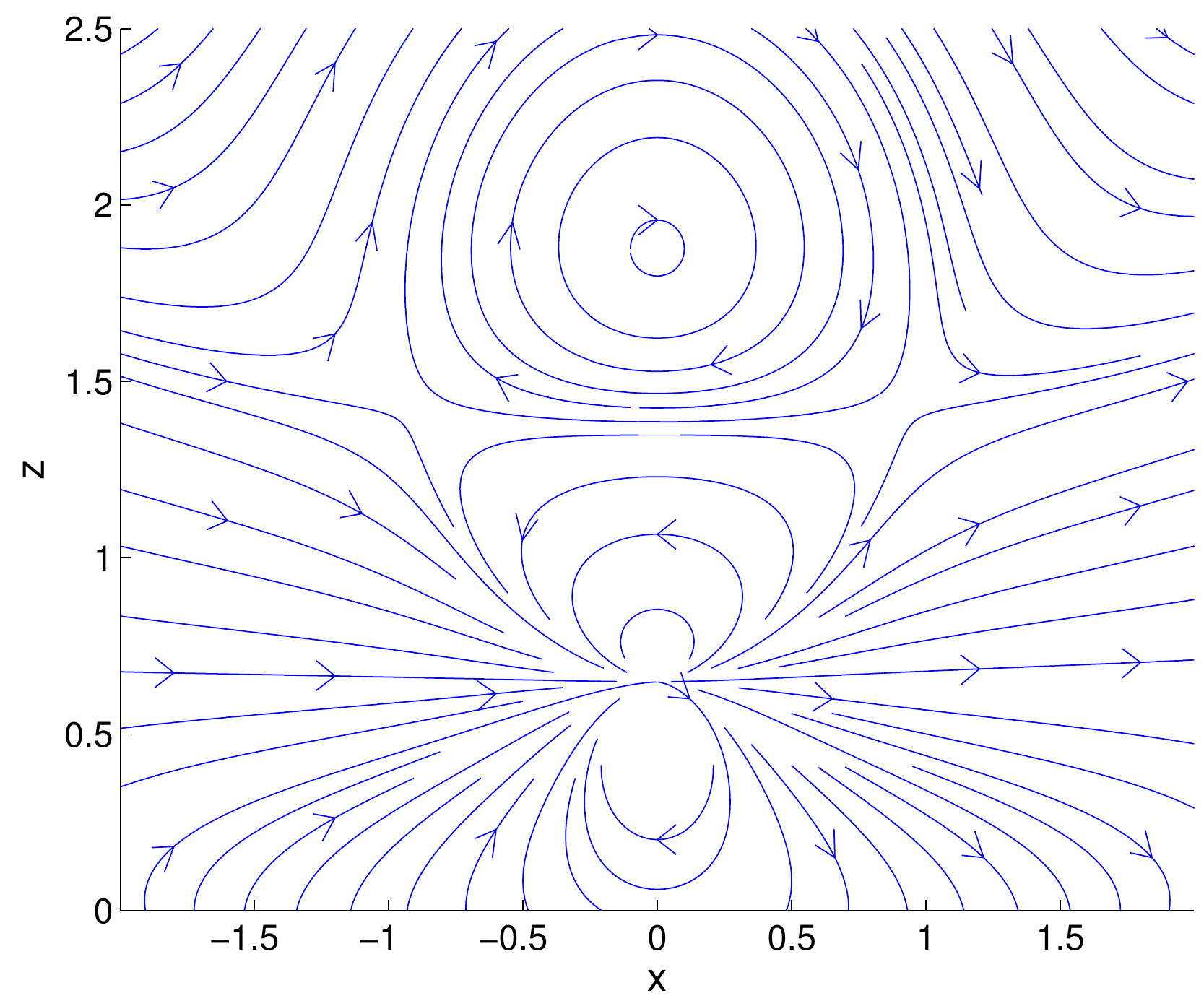}
\caption{Comparison of the magnetic field lines of a rotating dipole ($\omega=1$, $m=1$) as observed in the inertial reference frame ($g_H t=0$, left panel) and in the accelerated Rindler frame ($g_H t=1$, right panel). The dipole is located at $Z_s=1$.}
\label{mag_radiative}
\end{figure*}

And the magnetic field components are

	\begin{eqnarray}
	\label{Bx}B_{x}&=&\cosh(g_{H}\,t)\left[\frac{2\,\left(r\,\dot{m}_{x}+{m}_{x}\right)}{r^3}-\frac{\left(r^2-x^2\right)\left(r^2\,\ddot{m}_{x}+3\,r\,\dot{m}_{x}+3\,{m}_{x}\right)}{r^5}+\frac{x\,y\,\left(r^2\,\ddot{m}_{y}+3\,r\,\dot{m}_{y}+3\,{m}_{y}\right)}{r^5}\right]\nonumber \\
	&&+\sinh(g_{H}\,t)\left[\frac{\left(z\,\cosh(g_{H}t)-Z{s}\right)\left(\dot{m}_{x}+r\,\ddot{m}_{x}\right)}{r^3}\right],\\
	\label{By}B_{y}&=&\cosh(g_{H}\,t)\left[\frac{2\,\left(r\,\dot{m}_{y}+{m}_{y}\right)}{r^3}
	-\frac{\left(r^2-y^2\right)\left(r^2\,\ddot{m}_{y}+3\,r\,\dot{m}_{y}+3\,{m}_{y}\right)}{r^5}+\frac{x\,y\,\left(r^2\,\ddot{m}_{x}+3\,r\,\dot{m}_{x}+3\,{m}_{x}\right)}{r^5}\right]\nonumber \\
	&&+\sinh(g_{H}\,t)\left[\frac{\left(z\,\cosh(g_{H}t)-Z{s}\right)\left(\dot{m}_{y}+r\,\ddot{m}_{y}\right)}{r^3}\right], \\
	\label{Bz}B_{z}&=&\left(z\,\cosh(g_{H}t)-Z{s}\right)\left[\frac{r^2\,\left(x\,\ddot{m}_{x}+y\,\ddot{m}_{y}\right)+3\,\left(x\,{m}_{x}+y\,{m}_{y}\right)+3\,r\,\left(x\,\dot{m}_{x}+y\,\dot{m}_{y}\right)}{r^5}\right],
	\end{eqnarray}

where the spatial distance from the dipole $r$ is expressed in Rindler coordinates as follows
\begin{equation}
r=\sqrt{x^2+y^2+(z\,\cosh(g_{H}t)-Z{s})^2},
\end{equation} 
and the retarded proper time is 
\begin{equation}\label{proper time}
\tau_{*}=z\,\sinh(g_{H}t)-\sqrt{x^2+y^2+(z\,\cosh(g_{H}t)-Z{s})^2}\,\,.
\end{equation}
Since the magnetic dipole is static in Minkowski spacetime, the proper retarded time $\tau_{*}$ is the same as Minkowski retarded time $T_{*}$.
\end{widetext}

\clearpage

The magnetic field lines are calculated using the Runge–Kutta method to solve the system of ordinary differential equations
\begin{equation}
\frac{d\bm{r}}{ds}=\frac{\bm{B}}{|\bm{B}|},
\label{fline}
\end{equation}

where $ds$ is the element of the magnetic field line \cite{sarychev09}. 

In Fig.~\ref{3D_radiative} we compare the structure of the magnetic field (\ref{Bx})--(\ref{Bz}) in the Minkowski (left panel) and Rindler (right panel) spacetimes.
Time evolution of the field lines in the latter case is illustrated in Fig.~\ref{Rindler} for the two cases, $t=1$ (``early time'') and $t=3$ (``late tim''). In the Minkowski frame we would obtain a similarly shaped field line as shown for the Rindler spacetime in the early time (left panel of Fig.~\ref{Rindler}). However, as time passes, the structure of field lines in the Rindler frame evolves and gradually becomes more and more distorted and compressed due to the presence of the Rindler horizon as shown in the right panel of Fig. \ref{Rindler}.

In Fig.~\ref{el_radiative} we present the section of the electric field (\ref{Ex})--(\ref{Ez}) in the $(X,Z)$ [$(x,z)$, respectively] planes in the Minkowski (left panel, $g_H t=0$) and Rindler (right panel, $g_H t=1$) spacetimes. Corresponding magnetic field lines (\ref{Bx})--(\ref{Bz}) are compared in Fig.~\ref{mag_radiative}. We stress that these sections do not cover $y$-components of the fields which are generally nonzero as can be seen in the 3D view of magnetic field lines in Fig.~\ref{3D_radiative}.

\section{Electrovacuum model of a neutron star magnetosphere near SMBH}
\label{magnetosphere}
The electrodynamics of a rotating magnetized neutron star poses a difficult challenge which is being tackled by various approaches (see \cite{cerutti17} for a recent review of the topic). The force-free magnetodynamics of a neutron star in general relativity was previously studied by Komissarov \cite{komissarov11}, while the stationary vacuum solutions for electromagnetic fields around slowly rotating neutron stars were discussed in \cite{rezzolla01,petri13}. For the purpose of analyzing the effects of strong gravity onto the electromagnetic fields, the vacuum models become especially useful as they allow us to manifest the role of gravitation in the clear form. Therefore, for the exterior of the neutron star ($r>R$) we adopt a model of electrovacuum magnetosphere empty of any plasma or particles. On the other hand, the interior of a star ($r\leq R$) is supposed to be in a superconducting and superfluid state.

The magnetosphere is analyzed in the near-horizon limit, where the spatial curvature may be ignored and the metric may be approximated by that of Rindler. The electromagnetic field of a test charge in the near-horizon limit and its interaction with the horizon were studied in \cite{macdonald85}. More recently, the analytic solution for the magnetic dipole in the near-horizon limit was considered and the battery effect of the neutron star orbiting the SMBH was investigated \cite{orazio13}. The authors  analyzed the field of the magnetic dipole in uniform motion along the flat Rindler horizon. Here we focus on different astrophysically motivated aspects of the electrodynamics  of magnetized stars. In particular, we consider the rotating magnetic dipole which is arbitrarily inclined with respect to the rotation axis and study the structure of the resulting electromagnetic field.

It is well known that the induced electric field due to the rapid rotation of the magnetic dipole of a neutron star plays the main role in determining the characteristic of the pulsar magnetosphere. Because of its rotation, an electromotive field is induced such that the electric field in the corotating frame vanishes, $\bm{E}'=0$ \cite{conductor1,conductor2}.  From the transformation law between an inertial frame and a rotating frame we get
\begin{equation}
\bm{E}'=\bm{E}+(\bm{\omega} \wedge \bm{r}) \wedge \bm{B}
\end{equation}
where $\bm{r}$ is the position vector and $\bm{\omega}$ the rotation velocity vector of the star. We assume the presence of a pointlike magnetic dipole in its center. The external magnetic field in the quasistatic near zone for distances much less than the wavelength $\lambda={2\,\pi\,c}/{\omega}$ in Minkowski spacetime is \cite{conductor2}
\begin{widetext}
\begin{eqnarray}
B_{X}&=&\frac{m}{r^5}\left[\sin(\chi)\left\{3\,X\left[X\,\cos(\omega\tau)+Y\,\sin(\omega\tau)\right]-r^2\,\cos(\omega\tau)\right\}+3\,(Z-Z{s})\,X\,\cos(\chi)\right], \\
B_{Y}&=&\frac{m}{r^5}\left[\sin(\chi)\left\{3\,Y\left[X\,\cos(\omega\tau)+Y\,\sin(\omega\tau)\right]-r^2\,\sin(\omega\tau)\right\}+3\,(Z-Z{s})\,Y\,\cos(\chi)\right],\\
B_{Z}&=&\frac{m}{r^5}\left[3\,(Z-Z{s})\,\sin(\chi)\,\left[X\,\cos(\omega\tau)+Y\,\sin(\omega\tau)\right]+\cos(\chi)\left(3\,(Z-Z{s})^2-r^2\right)\right],
\end{eqnarray}
where the distance $r$ from the source is given in Minkowski coordinates as $r=\sqrt{X^2+Y^2+(Z-Z{s})^2}$.

Further we consider a star of radius $R$ rotating along the $Z$-axis with the inclination angle $\chi$ between the rotation axis and the magnetic moment. 
The external electric field in vacuum is quadrupolar and its components are

	\begin{eqnarray}
	E_{X}&=&-\frac{m\omega}{r^5}\left[(Z-Z{s})\sin(\chi)\left\{\frac{5X\,R^2}{r^2}\left(X\cos(\omega\tau)+Y\sin(\omega\tau)\right)+(r^2-R^2)\cos(\omega\tau)\right\}+R^2\,X\cos(\chi)\left(\frac{5(Z-Z{s})^2}{r^2}-1\right)\right], \nonumber \\
	\\
	E_{Y}&=&-\frac{m\omega}{r^5}\left[(Z-Z{s})\sin(\chi)\left\{\frac{5Y\,R^2}{r^2}\left(X\cos(\omega\tau)+Y\sin(\omega\tau)\right)+(r^2-R^2)\sin(\omega\tau)\right\}+R^2\,Y\cos(\chi)\left(\frac{5(Z-Z{s})^2}{r^2}-1\right)\right], \nonumber \\
	\\
	E_{Z}&=&\frac{m\,\omega}{r^5}\left[\sin(\chi)\left\{r^2+R^2-\frac{5\,(Z-Z{s})^2}{r^2}\right\}\,\left(X\cos(\omega\tau)+Y\sin(\omega\tau)\right)+R^2\,(Z-Z{s})\,\cos(\chi)\left(3-\frac{5\,(Z-Z{s})^2}{r^2}\right)\right].
	\end{eqnarray}
\end{widetext}
If we put $\chi=\frac{\pi}{2}$, $R=0$ (the case without a conductor) and neglect the time derivatives of $m$ (reducing to zero order in terms of $\omega$), we recover the electromagnetic fields in Minkowski given by (\ref{Bx-Mink})--(\ref{Bz-Mink}) and (\ref{Ex-Mink})--(\ref{Ez-Mink}), respectively. The electric field linear in $\omega$ vanishes.

We use the transformation relation from Minkowski to Rindler coordinates to obtain the electromagnetic fields in Rindler spacetime. From (\ref{Rindler transf}), (\ref{field-transf}) and (\ref{Rindler velocity}) we obtain the magnetic field components as 
\begin{widetext}
	\begin{eqnarray}
	B_{x}&=&\frac{m}{r^5}\left\{\cosh(g_{H}t)\Bigg[\sin(\chi)\left\{3\,x\left[x\,\cos(\omega\tau)+y\,\sin(\omega\tau)\right]-r^2\,\cos(\omega\tau)\right\}+3\,(z\,\cosh(g_{H}t)-Z{s})\,x\,\cos(\chi)\Bigg] \right.\nonumber \\
	&&-\omega\sinh(g_{H}t)\left[(z\,\cosh(g_{H}t)-Z{s})\sin(\chi)\left\{\frac{5\,R^2\,y}{r^2}\left(x\,\cos(\omega\tau)+y\,\sin(\omega\tau)\right)+(r^2-R^2)\,\sin(\omega\tau)\right\} \right. \nonumber \\
	&&\left. \left.+R^2\,y\,\cos(\chi)\left(\frac{5\,(z\,\cosh(g_{H}t)-Z{s})^2}{r^2}-1\right)\right] \right\}, \label{Magn-x-Cond-Rind}\\
	B_{y}&=&\frac{m}{r^5}\left\{\cosh(g_{H}t)\Bigg[\sin(\chi)\left\{3\,y\left[x\,\cos(\omega\tau)+y\,\sin(\omega\tau)\right]-r^2\,\sin(\omega\tau)\right\}+3\,(z\,\cosh(g_{H}t)-Z{s})\,y\,\cos(\chi)\Bigg] \right.\nonumber\\
	&&+\omega \sinh(g_{H}t) \left[(z\,\cosh(g_{H}t)-Z{s})\sin(\chi)\left\{\frac{5\,x\,R^2}{r^2}\left(x\,\cos(\omega\tau)+y\,\sin(\omega\tau)\right)+(r^2-R^2)\,\cos(\omega\tau)\right\}\right. \nonumber \\
	&&\left. \left.+R^2\,x\,\cos(\chi)\left(\frac{5\,(z\,\cosh(g_{H}t)-Z{s})^2}{r^2}-1\right)\right]\right\}, \label{Magn-y-Cond-Rind} \\
	B_{z}&=&\frac{m}{r^5}\Bigg[3\,(z\,\cosh(g_{H}t)-Z{s})\,\sin(\chi)\,\left[x\,\cos(\omega\tau)+y\,\sin(\omega\tau)\right]+\cos(\chi)\left(3\,(z\,\cosh(g_{H}t)-Z{s})^2-r^2\right)\Bigg], \label{Magn-z-Cond-Rind}
	\end{eqnarray}

where $\tau$ is defined by Eq.~(\ref{proper time}). The electric field components are 

	\begin{eqnarray}
	E_{x}&=&-\frac{m}{r^5}\left\{\omega\cosh(g_{H}t)\left[(z\,\cosh(g_{H}t)-Z{s})\sin(\chi)\left\{\frac{5\,R^2\,x}{r^2}\left(x\,\cos(\omega\tau)+y\,\sin(\omega\tau)\right)+(r^2-R^2)\,\cos(\omega\tau)\right\} \right. \right.\nonumber \\
	&&\left.+R^2\,x\,\cos(\chi)\left(\frac{5\,(z\,\cosh(g_{H}t)-Z{s})^2}{r^2}-1\right)\right]+\sinh(g_{H}t)\Bigg[\sin(\chi)\left\{3\,y\left[x\,\cos(\omega\tau)+y\,\sin(\omega\tau)\right]-r^2\,\sin(\omega\tau)\right\}  \nonumber \\
	&&\left.+3\,(z\,\cosh(g_{H}t)-Z{s})\,y\,\cos(\chi)\Bigg]\right\},\\
	E_{y}&=&-\frac{m}{r^5}\left\{\omega\cosh(g_{H}t)\left[(z\,\cosh(g_{H}t)-Z{s})\sin(\chi)\left\{\frac{5\,R^2\,y}{r^2}\left(x\,\cos(\omega\tau)+y\,\sin(\omega\tau)\right)+(r^2-R^2)\,\sin(\omega\tau)\right\} \right. \right. \nonumber\\
	&&\left.+R^2\,y\,\cos(\chi)\left(\frac{5\,(z\,\cosh(g_{H}t)-Z{s})^2}{r^2}-1\right)\right] -\sinh(g_{H}t)\Bigg[\sin(\chi)\left\{3\,x\left[x\,\cos(\omega\tau)+y\,\sin(\omega\tau)\right]-r^2\,\cos(\omega\tau)\right\} \nonumber \\  
	&&\left.+3\,(z\,\cosh(g_{H}t)-Z{s})\,x\,\cos(\chi)\Bigg]\right\}, \\
	E_{z}&=&\frac{m}{r^5}\Bigg[3\,(z\,\cosh(g_{H}t)-Z{s})\,\sin(\chi)\,\left[x\,\cos(\omega\tau)+y\,\sin(\omega\tau)\right]+\cos(\chi)\left(3\,(z\,\cosh(g_{H}t)-Z{s})^2-r^2\right)\Bigg].
	\end{eqnarray}
\end{widetext}

\section{Magnetic Null Points}
\label{nps}
The formation of magnetic null points (NPs) where all the components of the magnetic induction vector simultaneously vanish, i.e., $\bm{B}=(B_x,B_y,B_z)=(0,0,0)$ within the magnetosphere of a magnetic star, is an astrophysically important effect. Structure of the magnetic field lines around a NP is relevant for the processes of magnetic reconnection occurring in the presence of astrophysical plasma. Therefore we investigate whether the employed model of the magnetosphere supports the formation of NPs and how their presence and location depend on the parameters of the model.

\subsection{Analytical Method}
First we try to locate magnetic NPs analytically using some further assumptions to simplify long mathematical expressions. From (\ref{Magn-z-Cond-Rind}) we get

\begin{widetext}
	\begin{eqnarray}
	B_{z}=\frac{m}{r^5}\Bigg[3z_{R}\,\sin(\chi)\left[x\cos(\omega\tau)+y\sin(\omega\tau)\right]+\cos(\chi)\left(3\,z^2_{R}-r^2\right)\Bigg]=0, 
	\end{eqnarray}

where $z_{R}=z\,\cosh(g_{H}t)-Z{s}$ and $r\neq0$.
\end{widetext}

 From the above equation we obtain the following relation:  
\begin{equation}\label{Bz-zero}
\tan(\chi)=\frac{r^2-3\,z_{R}^2}{3\,z_{R}\,\left[x\cos(\omega\tau)+y\sin(\omega\tau)\right]}.
\end{equation}

Let us consider the axisymmetric case where the rotational axis and magnetic moment of the star have parallel orientation along the $z$-axis, i.e., $\chi=0$, which gives 
\begin{equation}\label{zero condition}
x^2+y^2-2\,z_{R}^2=0.
\end{equation}
Imposing this constrain in $B_{x}=0$ and $B_{y}=0$, we obtain
\begin{equation}\label{nuly}
x=0, \,\,\,\,\, y=0,
\end{equation}
and from (\ref{zero condition}) we get $z_{R}=0$, resulting in $r=0$, which means that in this case we have no magnetic NP with astrophysical significance. If we perform an analogical procedure for $\chi=\pi/2$, we obtain the same result. Thus for $\chi=0,\pi/2$ we have no NPs at finite $r$ outside $r=0$. Apparently, the inclination angle $\chi$ plays a key role in the discussion of NPs. However, for arbitrary $\chi$, the expressions become too complicated for the analytic treatment and therefore we proceed numerically. 

\subsection{Numerical Method}
Since we could not locate the NPs analytically in the general case, we switch to the numerical approach. The iterative root-finding method is used to search the relevant portion of the magnetosphere for the presence of NPs and to find their location with sufficient precision. 

We eventually confirm that NPs may develop in the employed model of the magnetosphere. The acceleration of the Rindler frame, which represents the gravitational pull of the nearby SMBH, appears crucial for the formation of NPs which are actually not allowed in the Minkowskian limit ($g_H =0$). We observe that magnetic nulls only emerge for the inclinations $\chi\neq 0, \pi/2$ which confirms previous analytical results. Rotation of the inclined dipole is also essential for the formation of the NPs; none of them is found if the rotation stops ($\omega=0$). Necessary conditions for the existence of the magnetic nulls in the given setup are thus (i) acceleration ($g_H>0$); (ii) inclination of the dipole $\chi\neq 0, \pi/2$; and (iii) rotation of the dipole $\omega>0$. 

\begin{figure}
	\center
	\includegraphics[scale=.5]{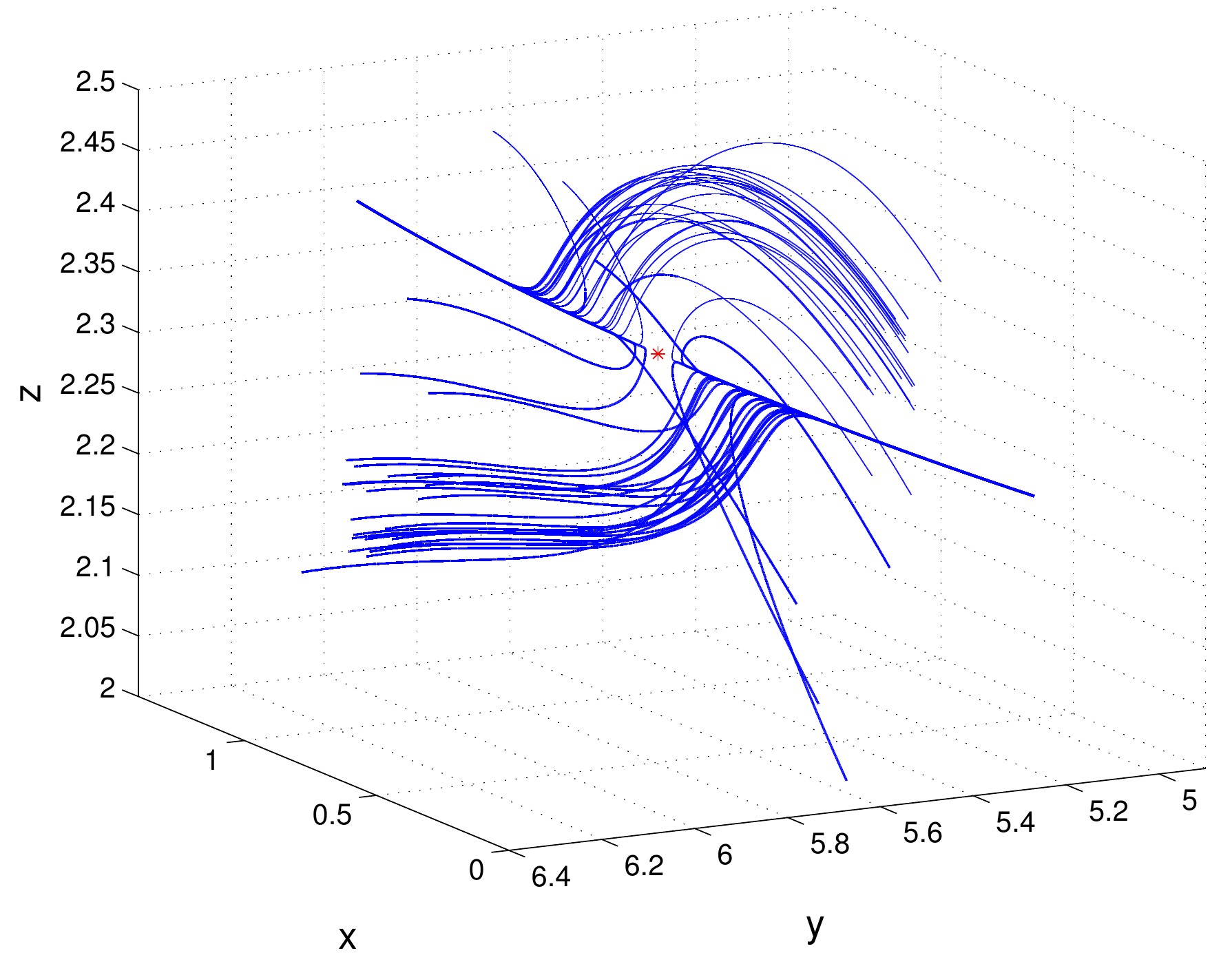}
	\caption{Structure of the magnetic field lines in the vicinity of the null point (red mark) located at $x_0=0.39$, $y_0=5.86$ and $z_0=2.35$. The following values of parameters are set: $Z_s =1$, $g_H=1$, $\omega=1$, $\chi=\pi/4$, $R=0$ and $t=1$.}
	\label{NP_3D}
\end{figure}

\begin{figure*}[htb]
	\center
	\includegraphics[scale=.37]{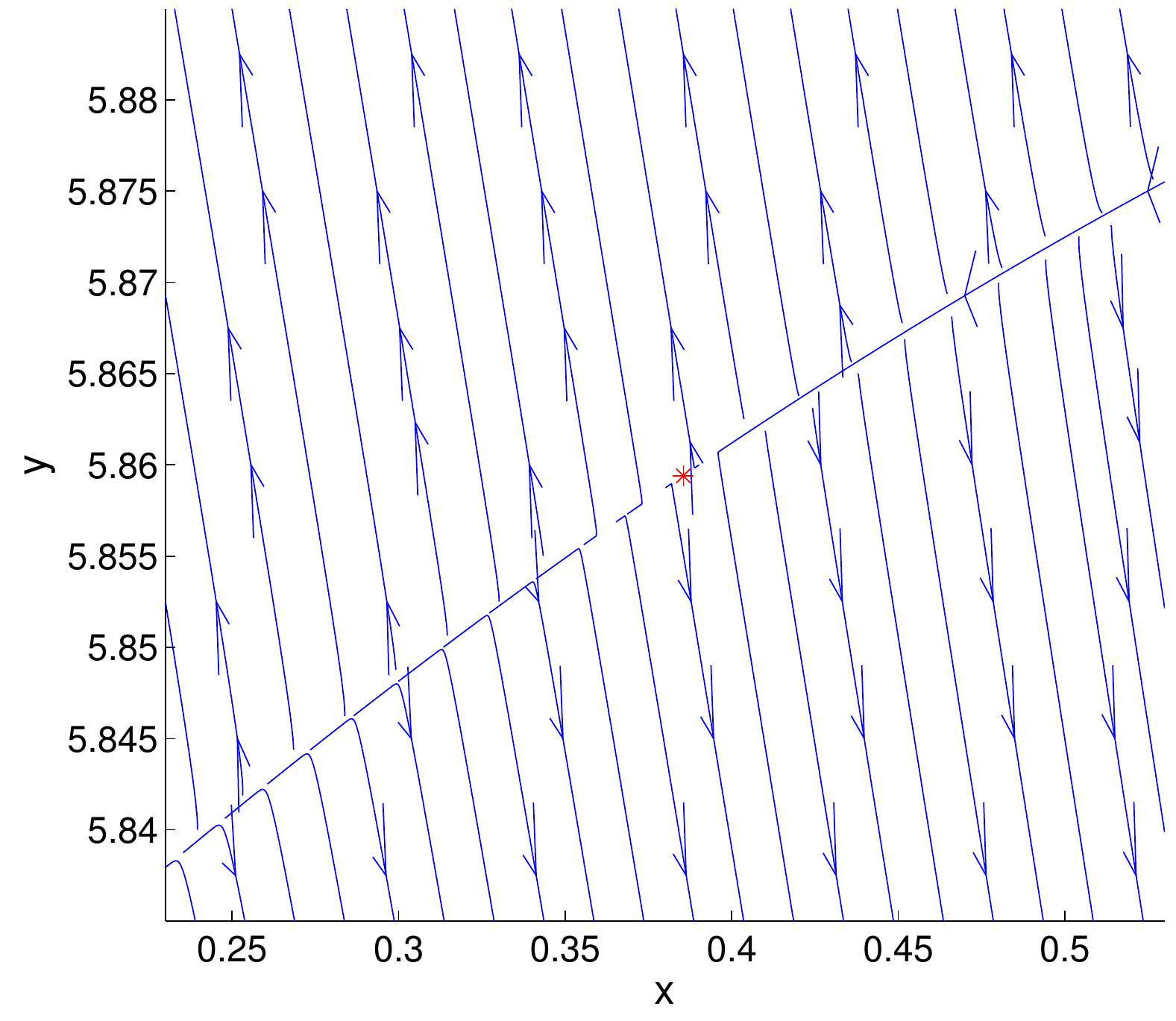}
	\includegraphics[scale=.37]{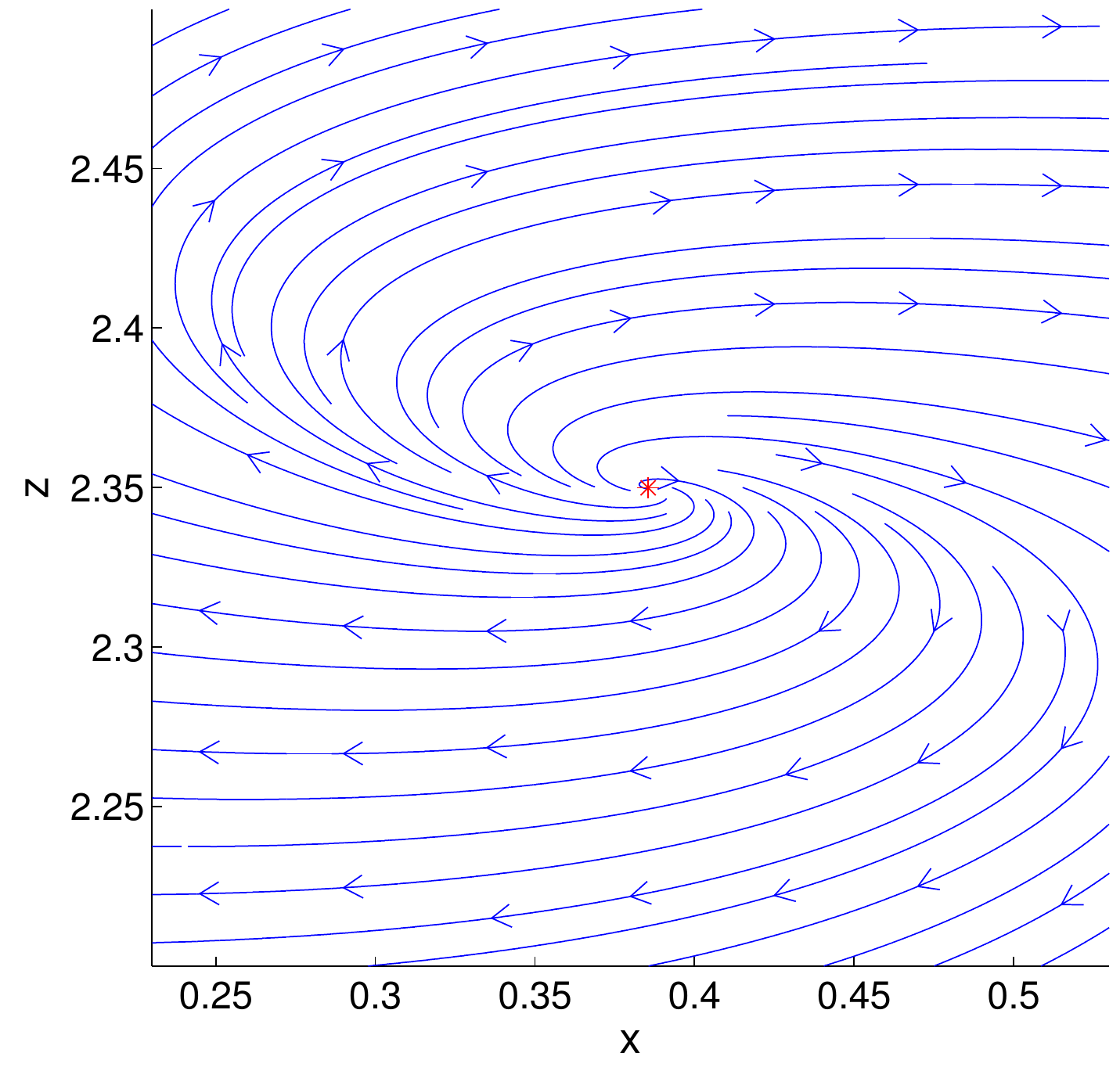}
	\includegraphics[scale=.37]{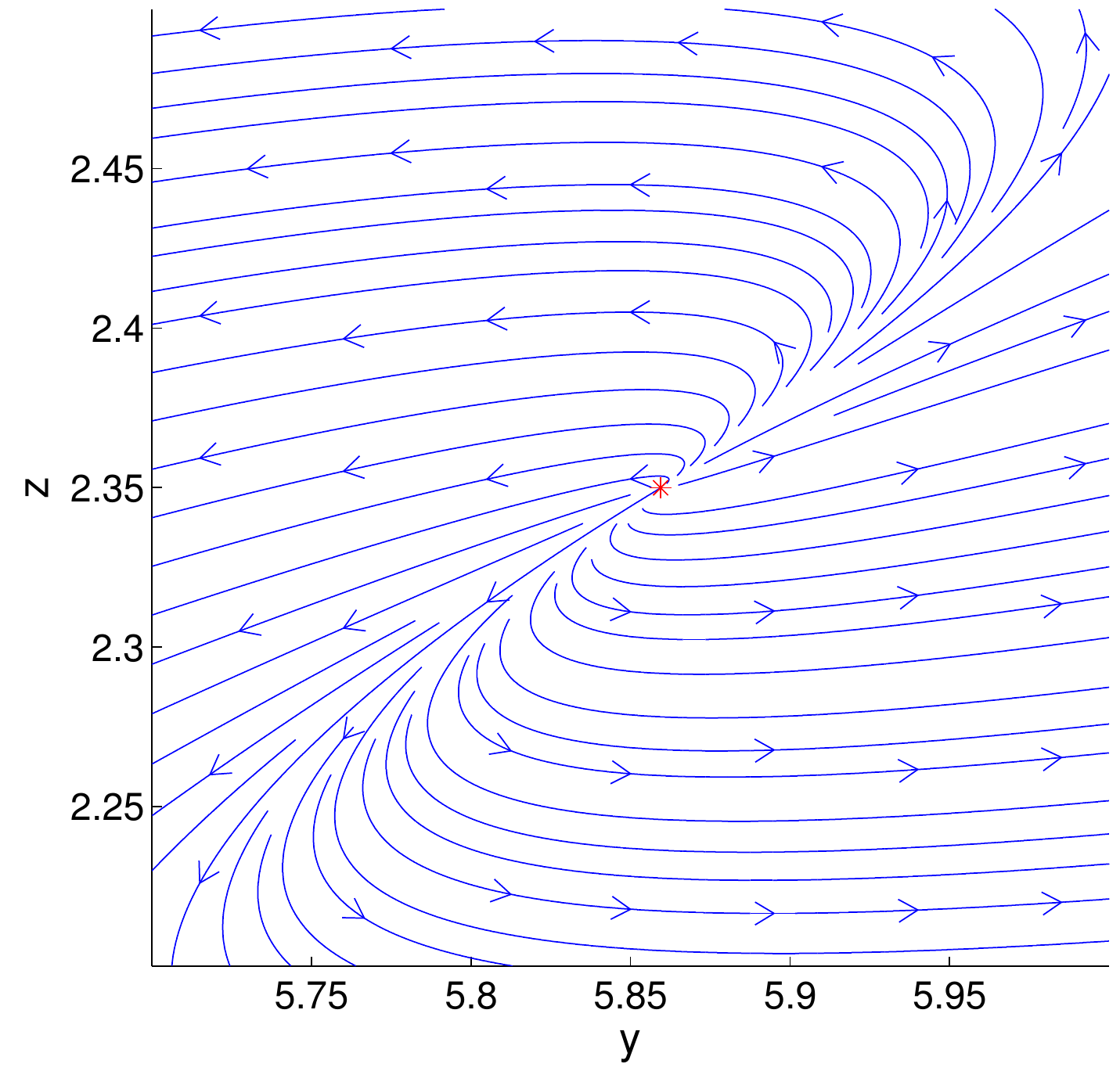}
	\caption{Two-dimensional sections of the magnetic field lines in the vicinity of the null point (red mark) located at $x_0=0.39$, $y_0=5.86$ and $z_0=2.35$. The same values of parameters as in Fig.~\ref{NP_3D} are used.}
	\label{NP_2D}
\end{figure*}
\begin{figure*}[htb]
	\center
	\includegraphics[scale=.3]{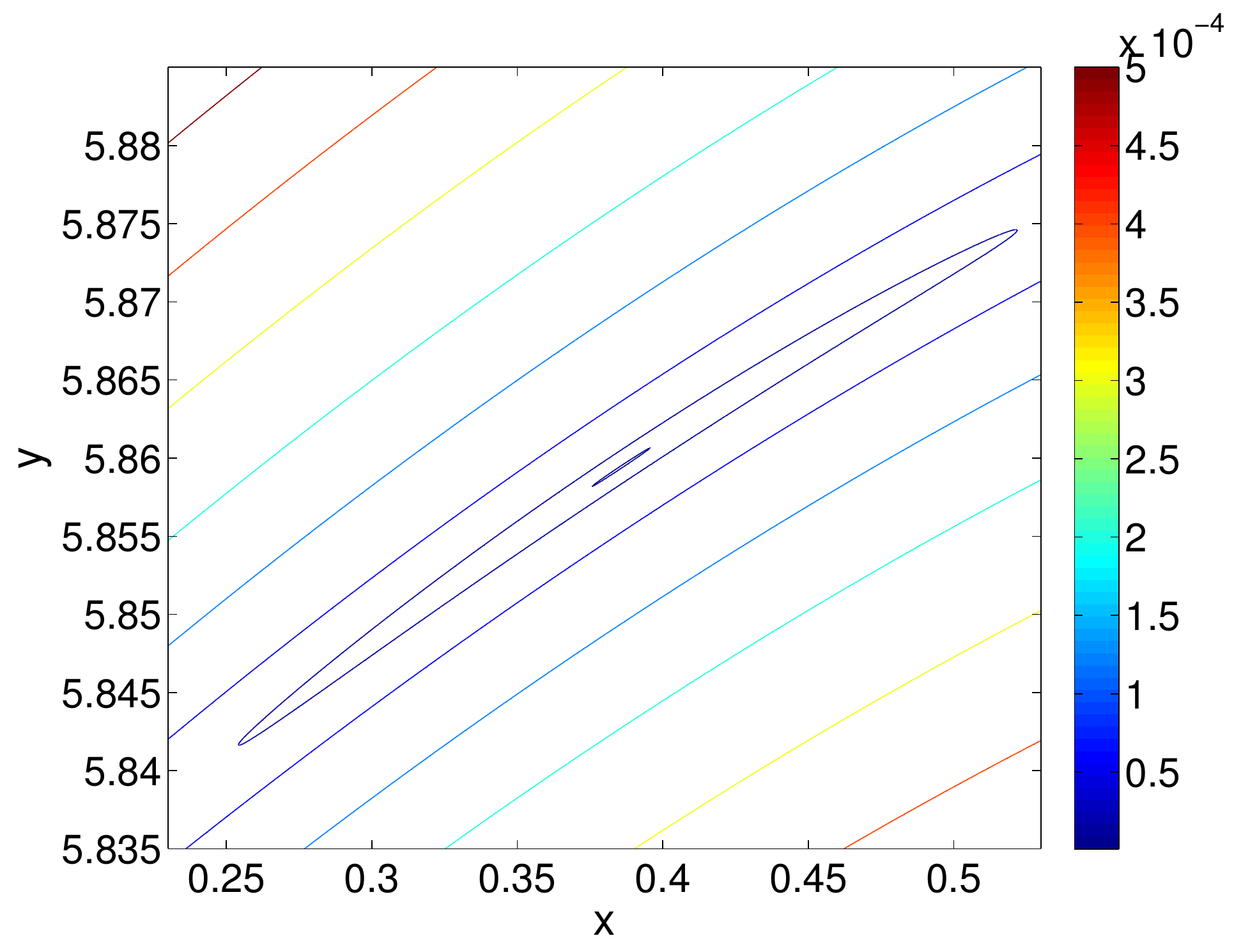}
	\includegraphics[scale=.3]{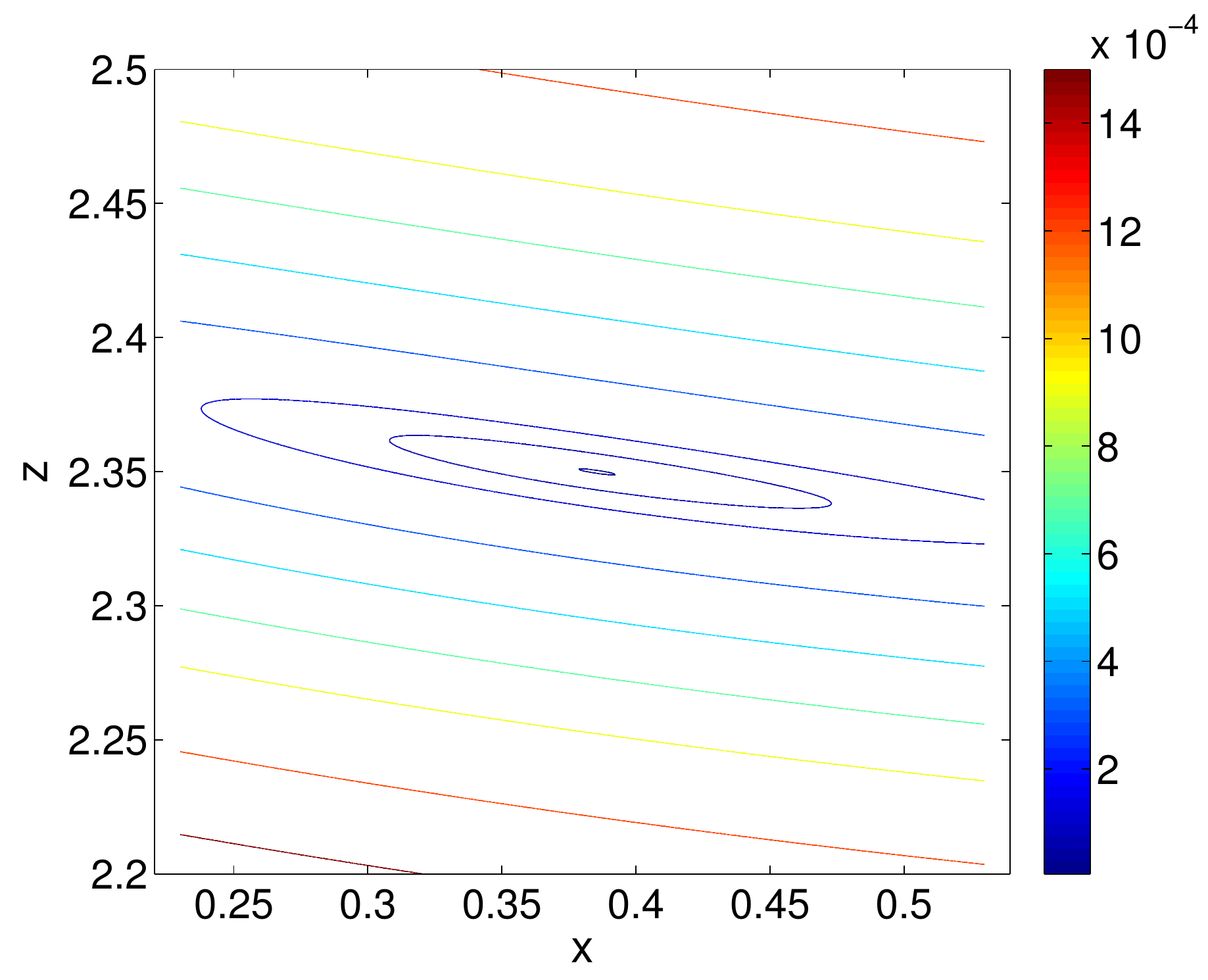}
	\includegraphics[scale=.3]{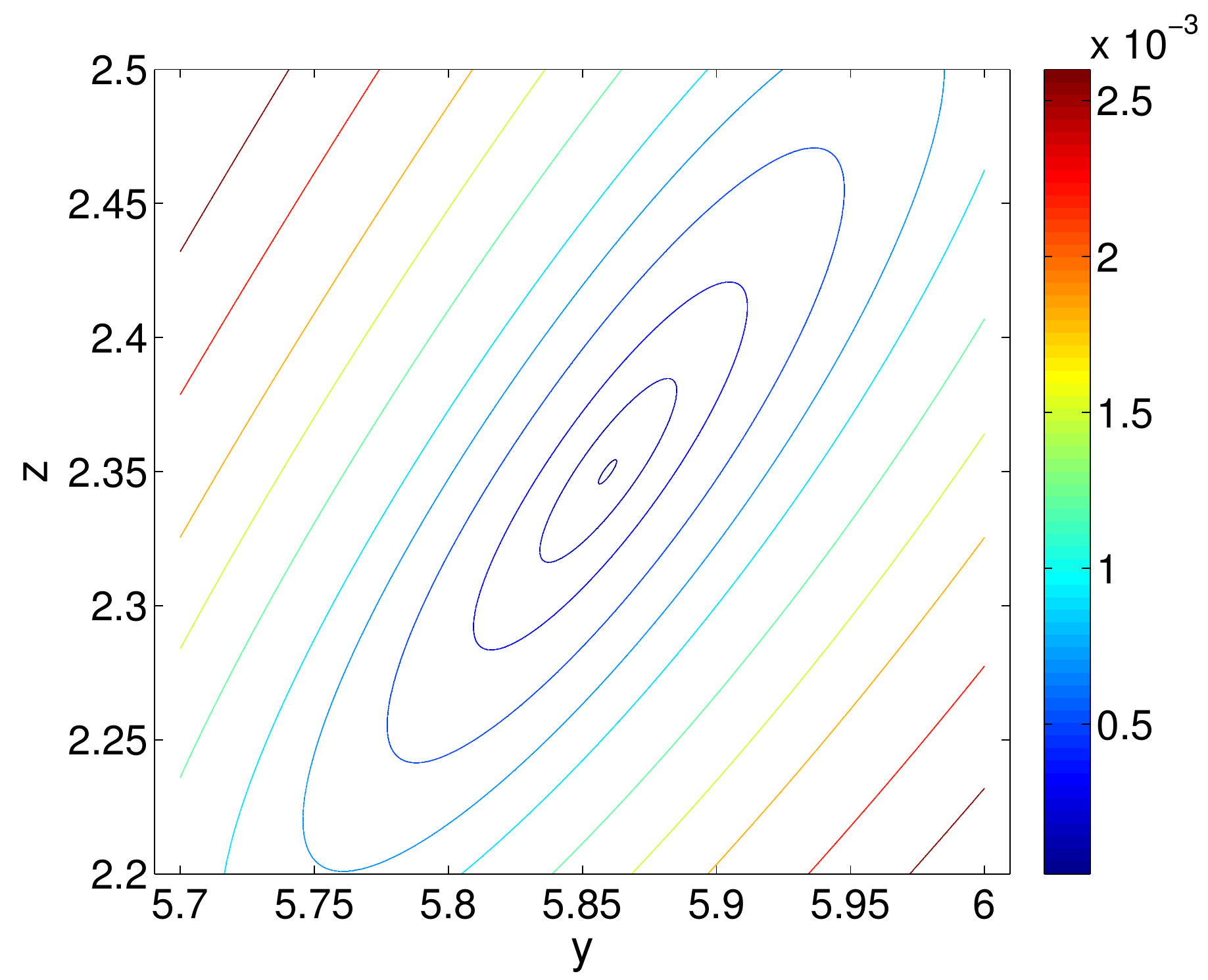}
	\caption{Isocontours of the magnetic field strength $B$ in the vicinity of the null point located at $x_0=0.39$, $y_0=5.86$ and $z_0=2.35$. The same values of parameters as in Fig.~\ref{NP_3D} and the same section planes as in Fig.~\ref{NP_2D} are used.}
	\label{NP_2D_contours}
\end{figure*}

In Fig.~\ref{NP_3D} we present a 3D view of a typical structure of magnetic field lines in the vicinity of the NP located for $\chi=\pi/4$, $Z_s=1$, $g_H=1$, $t=1$, $\omega=1$ and $R=0$. Two-dimensional sections of the same field structure in the planes $(x,y)$; $(x,z)$; and $(y,z)$ are shown in Fig.~\ref{NP_2D} and the corresponding isocontours of the magnetic field strength $B=(B_x^2+B_y^2+B_z^2)^{1/2}$ are shown in Fig.~\ref{NP_2D_contours}.

For the rest of the analysis we do not restrict ourselves to $z$-values corresponding to the vicinity of the horizon. We explore the behavior of the field in the larger domain and extend the search of NPs to regions where the Rindler metric does not necessarily approximate the Schwarzschild geometry.

\begin{figure}
	\center
	\includegraphics[scale=.45]{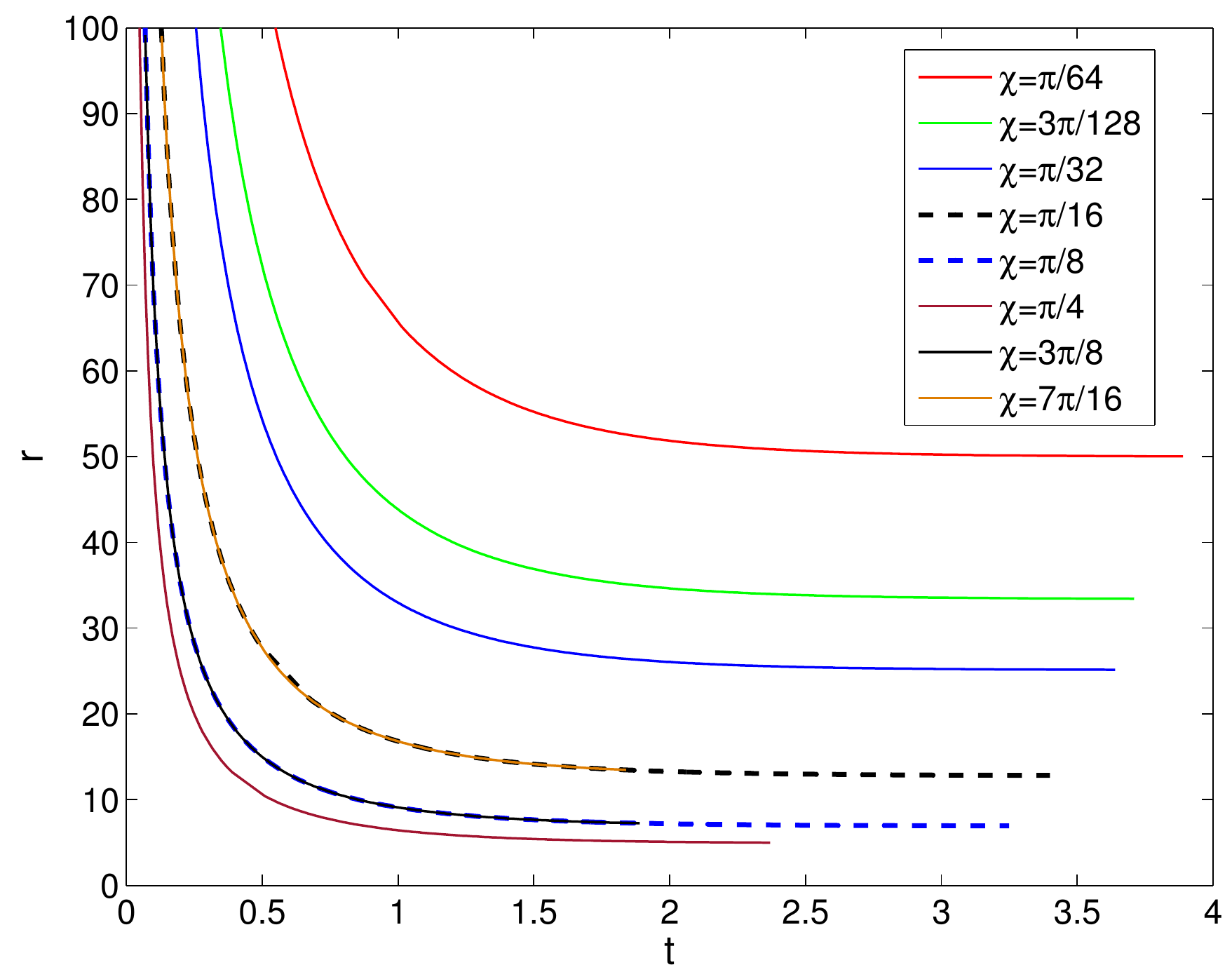}
	\caption{Distance of the magnetic null points from the dipole as a function of coordinate time $t$. Other parameters are fixed as $\omega=1$, $Z_s=1$, $g_H=1$ and $R=0$.}
	\label{NP_t}
\end{figure}

In the following we discuss the effect of the Rindler coordinate time $t$ which fundamentally affects the structure of the magnetic field and as a result it also impacts the formation of the NPs. For several values of the inclination angle $\chi$ we vary the coordinate time $t$ and seek for the NP while other parameters remain fixed as $\omega=1$, $Z_s=1$, $g_H=1$ and $R=0$. In Fig.~\ref{NP_t} we plot the NP's distance from the dipole $r$ as a function of $t$. For a small $t$ the NPs generally emerge far away from the dipole; then they become closer and the distance remains almost constant for some period of time after which the NPs vanish. This period decreases as the inclination grows; i.e., higher inclination makes the NP vanish earlier. Increasing the inclination we observe that the distance between the dipole and NP decreases until the value $\chi=\pi/4$ is reached. For higher inclinations this trend inverts and the distance starts to grow. Moreover, we observe that the curve for $\chi=3\pi/8$  coincides with the one for $\chi=\pi/8$ and similarly the curves for $\chi=7\pi/16$ and $\chi=\pi/16$ coincide, which suggests that the distance of NPs follows some kind of symmetry around $\chi=\pi/4$. For all tested values of $\chi$ we were able to locate null points also in the close vicinity of the horizon (with $z<1$) where Rindler spacetime faithfully approximates the Schwarzschild metric.

\begin{figure}[ht]
	\center
	\includegraphics[scale=.45]{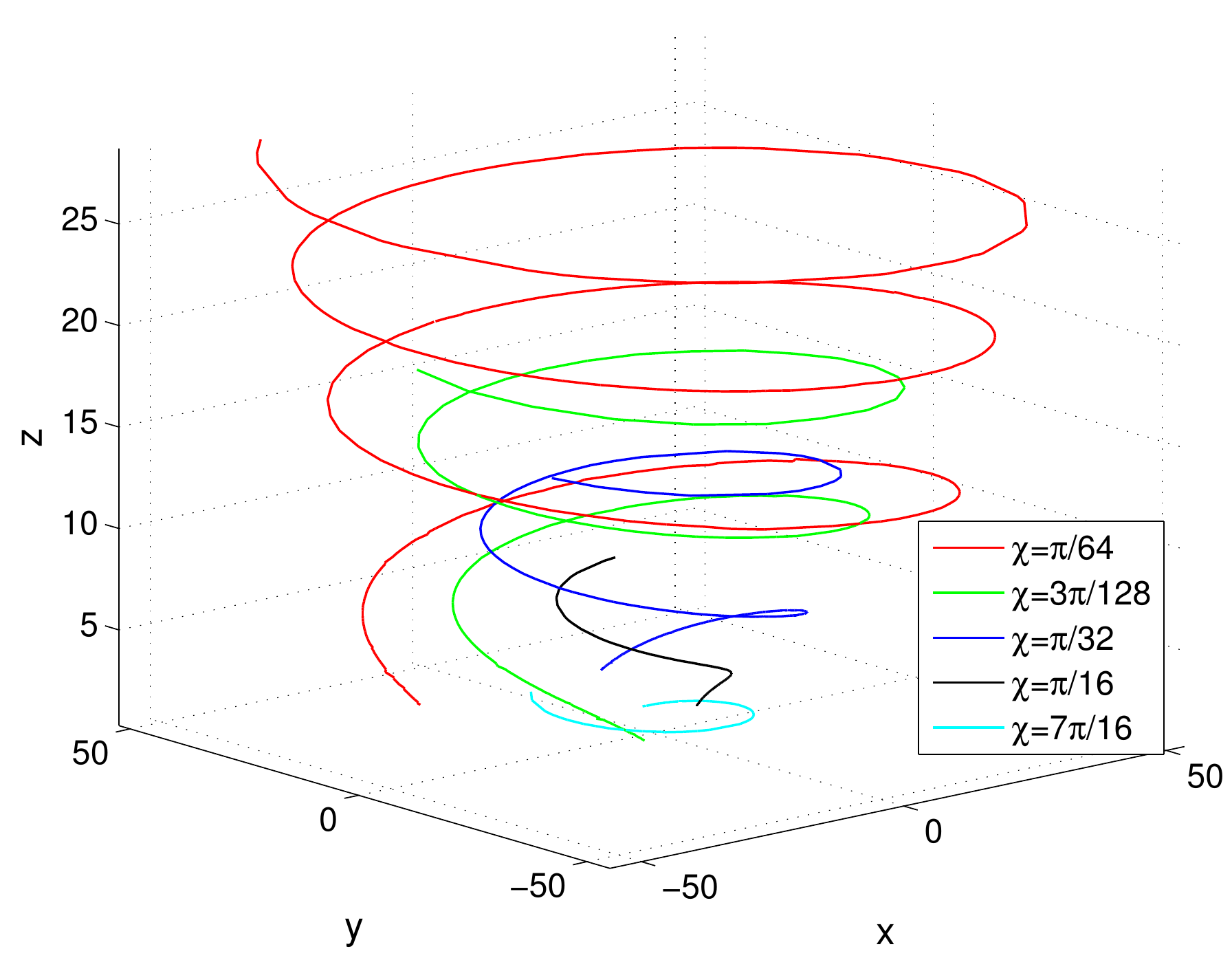}
	\caption{Location of the magnetic null points with Rindler time varying  in the interval $t\in\left<0.9,2.9\right>$. The same values of parameters as in Fig.~\ref{NP_t} are used here.}
	\label{NP_xyz}
\end{figure}

Location of the subset of NPs presented in Fig.~\ref{NP_t} is shown in a 3D view in Fig.~\ref{NP_xyz}. This shows that the above-mentioned symmetry only appears when we discuss the distance $r$ of the NPs, while their actual locations differ. Rotation of the dipole is parametrized by the proper time $\tau$ with the frequency $\omega$. The resulting electromagnetic field is $\tau$-periodic with the period $2\pi/\omega$. However, as $t$ increases, the distance of the NP from the dipole generally decreases as shown in Fig.~\ref{NP_t}. Combined with the simultaneous evolution of the proper time $\tau$, the location of the NP follows a helical trajectory with decreasing radius as $t$ grows (Fig.~\ref{NP_xyz}).

\begin{figure}[ht]
	\center
	\includegraphics[scale=.48]{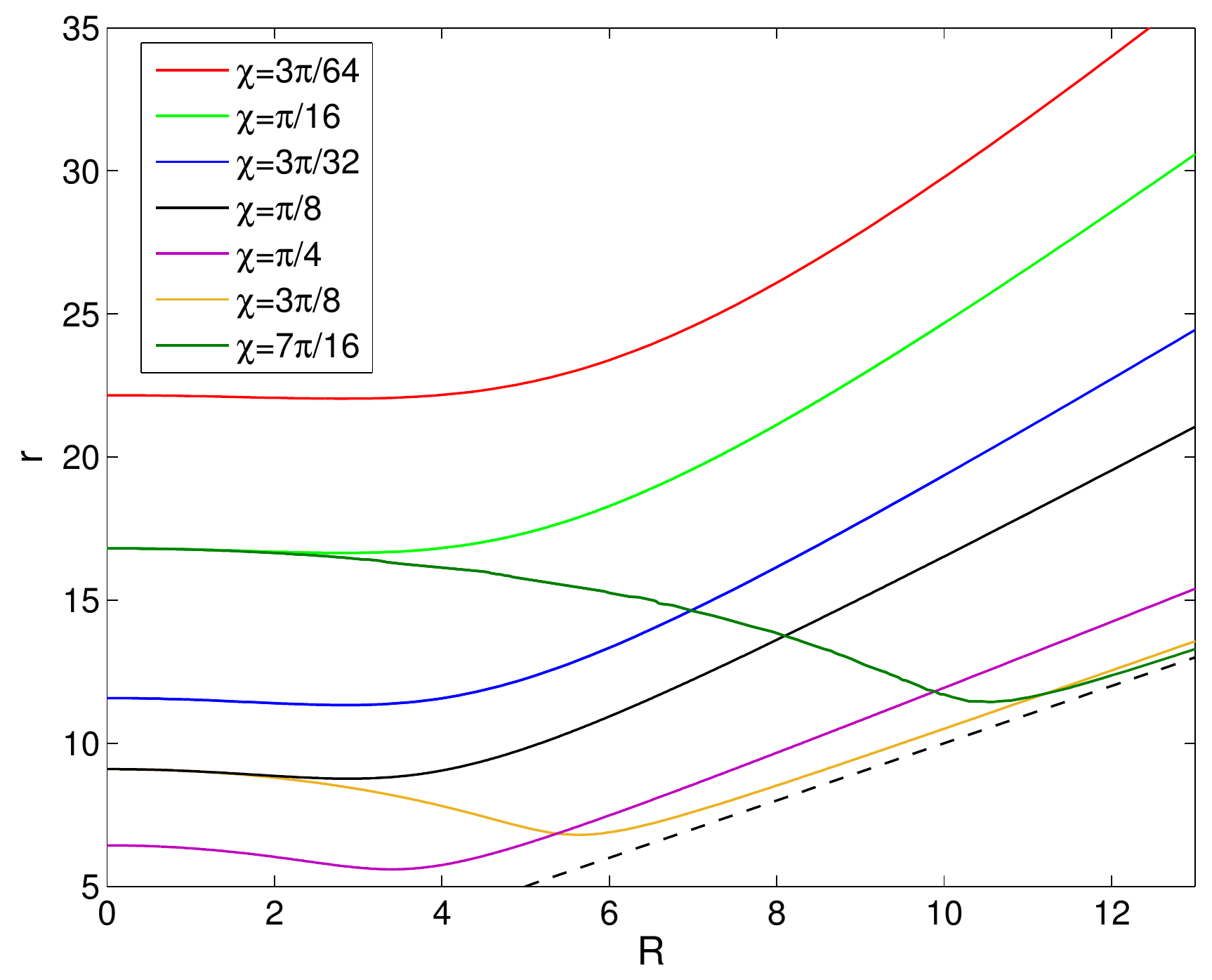}
	\caption{Distance of the magnetic NPs from the dipole as a function of the radius of the conductor $R$. Values of the parameters are fixed as $\omega=1$, $Z_s=1$, $g_H=1$ and $t=1$. The black dashed line shows the outer boundary of the conductor (surface of the star).}
	\label{NP_R}
\end{figure}

So far, we have considered the medium as nonconducting, i.e., the radius of the  conducting zone was set as $R=0$ in the discussion. The effect of conductivity on the structure of the field is important and the formation of the NPs is also  strongly affected when $R>0$ is considered. In Fig.~\ref{NP_R} we plot the distance of the NPs from the dipole as a function of $R$. At first the distance of the NPs slightly decreases as $R$ rises. The drop is almost negligible for small inclinations; however, for more inclined dipoles it becomes evident. Nevertheless, as $R$ further rises it starts to push NPs farther from the source and the boundary of the neutron star (black dashed line in Fig.~\ref{NP_R}). We observe that in the case of inclinations $\chi>\pi/4$ the NPs may get very close to the surface of the star as $R$ increases. Nevertheless, they never cross it, which suggests that in a given model the NPs cannot form within the superconducting interior of the neutron star.  

\section{Conclusions and final remarks}
\label{conclusions}
In this paper we studied a simplified vacuum model of a magnetosphere of a rapidly rotating neutron star (or magnetar) near a supermassive black hole represented by an inclined rotating magnetic dipole in the reference frame of an accelerated observer (Rindler spacetime). We investigated the structure of the electromagnetic field and, in particular, we focused on the formation of magnetic null points as a remarkable feature of the field topology with important consequences for the dynamics of astrophysical plasma.

We observed that magnetic null points may emerge in a given setup. The null points develop only if the inclination angle $\chi\neq 0, \pi/2$ (i.e., in the axisymmetric case the NPs do not appear). Rotation of the dipole and the acceleration of the observer are both necessary for the formation of NPs. We have numerically located the NPs for different values of $\chi$ and discussed their location as a function of coordinate time $t$ and radius of the conductor $R$. We conclude that NPs only form for some period of coordinate time $t$ and that they are always located outside the conducting region of a magnetic star.

Previously, we have also studied the formation of magnetic NPs in the curved spacetime of a rotating Kerr black hole \cite{karas14,karas12,karas09}. Namely, we investigated the near-horizon structure of an asymptotically uniform magnetic field (aligned or inclined with respect to the axis of rotation). The black hole was supposed to drift through the field with constant velocity. In the resulting field measured by a zero angular momentum observer (ZAMO) comoving with the black hole, the null points of the magnetic field developed only for nonzero values of spin and for sufficiently high drift velocity. Moreover, the nonaxisymmetry was also necessary for the formation of NPs (i.e., no NPs were found for the case where the spin, the magnetic field and the boost direction were all aligned). Comparing these results with the present discussion of NPs in the flat spacetime, we confirm the essential role of rotation and nonaxisymmetry. However, the fundamental effect of frame-dragging, which was necessary for the formation of NPs near the Kerr black hole, does not operate in the flat spacetime. Nevertheless, the acceleration of the Rindler observer is able to mimic this effect of curved spacetime and supports the formation of magnetic null points even in the case of flat spacetime.  

The analyzed model does not aim to provide a realistic description of the magnetosphere of a neutron star in which the role of charged matter and currents cannot be neglected. Nevertheless, we were basically interested in the gravitational effects on the structure of the electromagnetic fields and therefore we considered an electrovacuum solution. In our previous papers \cite{karas12,karas09}, we have shown that the curved geometry of a rotating black hole may lead to the formation of X-type null points  in the asymptotically uniform vacuum magnetic field. Here we raised and answered the question of whether and under which circumstances  the null points could appear even without curvature in the flat Rindler spacetime approximating the Schwarzschild geometry close to the horizon of the static black hole. 

Although the full description of the astrophysical process of magnetic reconnection is obviously beyond the scope of a vacuum model, we suggest that strong gravity effects analyzed in this paper might support the reconnection process as they entangle the field lines in a suitable way. We show that gravitation in an electrovacuum field may even create such a field topology which corresponds to the magnetic reconnection operating in the presence of charges and currents. These results are mostly of a theoretical interest and cannot be directly applied for the description of actual magnetospheres. Nevertheless, they clearly illustrate the complexity of relativistic effects which are relevant for the physics of compact objects.

\begin{acknowledgments}
This work was supported from the following grants of the Grant agency of the Czech republic: No. 17-06962Y (O. K.), No. 17-13525S (T. T.) and No. 17-16287S (V. K.). We would like to thank T. Ledvinka and O. Sv\'{i}tek for numerous discussions. 
\end{acknowledgments}

\end{document}